\documentclass[aps,pra,reprint,showpacs,groupedaddress]{revtex4-1}

\usepackage[utf8]{inputenc}
\usepackage{amsmath,amssymb,amsfonts}
\usepackage{graphicx}
\usepackage{psfrag}

\usepackage{color}

\usepackage{hyperref}

\begin{document}

\title{A two-dimensional quantum gas in a magnetic trap}

\author{K Merloti, R Dubessy, L Longchambon, A Perrin, P-E Pottie\footnote{Present address: LNE-SYRTE, Observatoire de Paris, CNRS, UPMC, F-75014 Paris, France}, V Lorent and H Perrin}

\address{Laboratoire de physique des lasers, CNRS, Universit\'e Paris 13, Sorbonne Paris Cit\'e, 99 avenue J.-B. Cl\'ement, F-93430 Villetaneuse}

\begin{abstract}
We present the first experimental realization of a two-dimensional quantum gas in a purely magnetic trap dressed by a radio frequency field in the presence of gravity. The resulting potential is extremely smooth and very close to harmonic in the two-dimensional plane of confinement. We fully characterize the trap and demonstrate the confinement of a quantum gas to two dimensions. The trap geometry can be modified to a large extent, in particular in a dynamical way. Taking advantage of this possibility, we study the monopole and the quadrupole modes of a two-dimensional Bose gas.
\end{abstract}


%\submitto{\NJP}

\maketitle

\section{Introduction}
For the last ten years, dramatic progress has been made in the investigation of low dimensional quantum gases~\cite{Houches2003}. When quantum gases are confined to one or two dimensions, the role of quantum correlations is strongly enhanced and the physics changes qualitatively~\cite{Bloch2008RMP}. For example, one-dimensional bosons behave as fermions in the Tonks-Girardeau regime~\cite{Tonks1936,Girardeau1960}, or a homogeneous bosonic gas restricted to two dimensions undergoes a specific transition to a superfluid state described by the Berezinskii-Kosterlitz-Thouless theory (BKT)~\cite{Berezinskii1972,Kosterlitz1973,Prokofev2001}, below a critical temperature.

While one-dimensional quantum gases have been investigated experimentally some time ago~\cite{Laburthe2004a,Kinoshita2004,Paredes2004a}, experimental results have been obtained more recently with two-dimensional quantum gases, including the observation of the BKT phase~\cite{Hadzibabic2006,Clade2009a} and the measurement of the equation of state~\cite{Rath2010}. In many respects, the dimension two is a critical dimension. Bose-Einstein condensation does not occur in a homogeneous two-dimensional gas, unless at strictly zero temperature~\cite{Petrov2004}. Long range order is destroyed, but the correlation function decreases algebraically which preserves a partial coherence on finite size systems~\cite{Bloch2008RMP,Plisson2011}. Moreover, interacting two-dimensional gases in a harmonic trap present a scale invariance~\cite{Hung2011} which is at the origin of the absence of damping of the breathing mode~\cite{Pitaevskii1997,Chevy2002}. Very recently, the influence of disorder on the phase coherence of a two-dimensional Bose gas has been investigated experimentally~\cite{Allard2012,Beeler2012}.

In order to reach the quasi two-dimensional regime for the atomic quantum gas,  all the relevant energies -- chemical potential $\mu$, temperature $k_BT$ -- must be lower than the transverse confining energy $\hbar\omega_z$ along the strongly confined direction $z$.
All the experiments which achieved this regime~\cite{Hadzibabic2006,Clade2009a,Plisson2011,Hung2011,Goerlitz2001,Rychtarik2004} made use of dipole traps~\cite{Grimm2000} to strongly confine the atoms in one direction.
The drawback of dipole traps is that, even if spontaneous photon scattering can be avoided by the use of high power and large detunings~\cite{Grimm2000}, pointing stability remains an issue for long time scale experiments~\cite{Gehm1998}. Moreover, light scattered by the windows of the vacuum chamber or any other surface close to the atoms may lead to interferences with the main light field and induce local defects in the trapping potential~\cite{Rychtarik2004}.
These defects are detrimental to experiments requiring a high smoothness of the trapping potential, or a high degree of control on the application of a tailored disorder~\cite{Allard2012,Beeler2012}.
On the other hand, macroscopic magnetic traps give access to smoothly varying potentials.
In this paper, we demonstrate the confinement of a quantum gas to two dimensions, using only magnetic and radio frequency (rf) fields~\cite{Zobay2001}.
Both fields are generated by macroscopic coils of centimetre scale and placed at a distance of order one centimetre from the atomic cloud.
In this way, any high frequency spatial noise is filtered out and the fields seen by the atoms vary smoothly, in a very well controlled way.
Moreover, the trap geometry can easily be dynamically controlled by the choice of the magnetic field gradient or the rf field amplitude, frequency and polarization.
Finally, we demonstrate long lifetimes and negligible heating rates in the trap.
This makes this setup very promising for the study of collective excitations or experiments involving a controlled, additional, disordered potential.

The paper is organized as follows: the anisotropic adiabatic trapping potential and its main characteristics are presented in section~\ref{sec:trap}.
Section~\ref{sec:BEC} is devoted to the experimental implementation of the trap.
In section~\ref{sec:2D} we show how to reach the quasi two-dimensional regime for the quantum gas with our setup.
In this regime, two collective modes of a two-dimensional gas, the quadrupole and the monopole modes, are evidenced, as demonstrated in section~\ref{sec:modes}.

\section{A very anisotropic dressed magnetic trap}
\label{sec:trap}
In this section, we recall the characteristics of adiabatic potentials~\cite{Zobay2001} and give the key features of the rf-dressed quadrupole trap discussed in this paper.

\begin{figure}[t]%
	\centering
 	\includegraphics[width=0.73\linewidth]{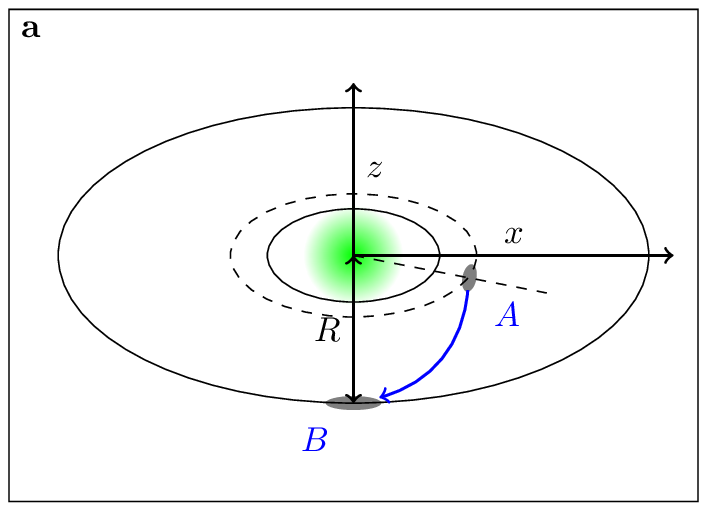}\hspace{0.02\linewidth}\includegraphics[width=0.25\linewidth]{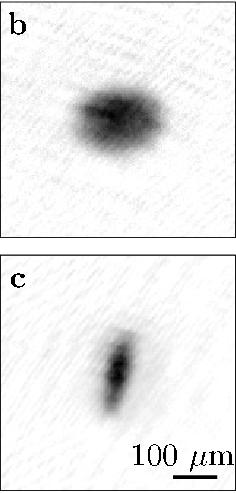}%
	\caption{(colour online)
	(a): Sketch of the rf-dressed trap loading procedure.
	Starting from a plugged quadrupole trap~\cite{Dubessy2012a} with the atoms trapped near $A$, we switch on an rf dressing field at a frequency below resonance (inner ellipsoid) and slowly ramp up the rf frequency to catch the atoms when it crosses resonance (dashed ellipsoid).
	If the rf frequency is increased further, the isomagnetic ellipsoid radius increases as well and the atoms don't feel the plug beam any longer. They fall to the bottom $B$, at a distance $R$ from the centre given by Eq.~\eqref{eq:eqpos}.
	The grey shaded areas indicate schematically the atomic cloud in the initial and final traps (not to scale). (b) and (c): Absorption imaging of the condensate after a 25-ms time-of-flight. The probe beam propagates along $y$. (b): Atoms released from the initial plugged trap at $A$. (c): Atoms released from the dressed quadrupole trap at $B$. The anisotropy is reversed, the final trap being more compressed in the vertical direction.
	}
		\label{fig:trap}
\end{figure}

The anisotropic magnetic trap is a `dressed trap', resulting from the adiabatic potential experienced by atoms placed in a magnetic quadrupole field and dressed by an rf field~\cite{Zobay2001,Morizot2007}. 
The static quadrupole field $\mathbf{B}_0(\mathbf{r}) = b'(x\,\mathbf{e}_x + y\,\mathbf{e}_y - 2 z\,\mathbf{e}_z)$, with a radial gradient $b'$, is produced by a pair of coils of vertical axis $z$.
The isomagnetic surfaces of the quadrupole field are defined by $r_e(\textbf{r})=r_0$ where the effective radius $r_e$ is
\begin{equation}
r_e(\textbf{r}) = \sqrt{x^2+y^2+4z^2}.
\label{eq:ellipsoid}
\end{equation}
As depicted in figure~\ref{fig:trap}(a), these surfaces  are ellipsoids of symmetry axis $z$, and semi-axes of length $r_0$ and $r_0/2$ along the horizontal and vertical directions, respectively.
The corresponding Zeeman splitting in the ground state of spin $F$ is
\begin{equation}
\hbar \alpha r_e(\textbf{r})
\end{equation}
where we have introduced the gradient in frequency units $\alpha=|g_F|\mu_B b'/\hbar$, $g_F$ being the Land\'e factor and $\mu_B$ the Bohr magneton.

The strongly anisotropic trap for the two-dimensional gas is based on this static inhomogeneous magnetic field, combined with an rf field~\cite{Zobay2001} of frequency $\omega_{\rm rf}$, circularly polarised along $z$: $\mathbf{B}_{\rm rf}(\mathbf{r},t) = B_1\left[\cos(\omega_{\rm rf}t)\,\mathbf{e}_x + \sin(\omega_{\rm rf} t)\,\mathbf{e}_y\right]$. 
Due to the position dependent orientation of the static magnetic field $\mathbf{B}_0(\mathbf{r})$, the effective Rabi coupling $\Omega(\textbf{r})$ between the Zeeman sub-states depends on the position $\mathbf{r}$ in the following way:
\begin{equation}
\Omega(\textbf{r}) = \frac{\Omega_0}{2}\left[1-\frac{2z}{r_e(\textbf{r})}\right],
\label{eq:Rabi}
\end{equation}
where $\hbar\Omega_0 = |g_F| \mu_B B_1$.
We note that the coupling, on a given isomagnetic surface, is maximum at the bottom of the ellipsoid and cancels at the top, where non-adiabatic spin flips can occur.

Assuming that $\omega_{\rm rf} \gg \Omega_0$ such that the rotating wave approximation can be applied, and that the atoms follow adiabatically a given dressed state of the atom and rf field system, the total adiabatic potential experienced by the dressed atoms in the sub-state $F,m_F$ in the presence of gravity is then given by~\cite{Morizot2007}: 
\begin{equation}
 V_{m_F}(\textbf{r})= m_F\hbar\sqrt{[\alpha r_e(\textbf{r})-\omega_{\rm rf}]^2+\Omega(\textbf{r})^2} + Mgz
 \label{eq:V}
\end{equation}
where $M$ is the atomic mass and $g$ the gravitational acceleration.

An avoided crossing is created in the region where the radio frequency field is resonant with the energy difference between the Zeeman states, that is on the isomagnetic surface $r_e(\textbf{r})=r_0$ where $r_0=\omega_{\rm rf}/\alpha$.
The adiabatic potential given at Eq.~\eqref{eq:V} confines the Zeeman states with the right spin orientation close to this isomagnetic surface. We chose the direction of the local quantization axis in such a way that the trapped states are the sub-states with positive $m_F$.
The atoms with $m_F>0$ will be trapped provided that the vertical magnetic field gradient exceeds gravity, namely $\varepsilon < 1$ where $\varepsilon$ is defined as $\varepsilon = Mg/(2m_F\hbar\alpha)$.
On the other hand, the vertical gradient of Rabi coupling, see Eq.~\eqref{eq:Rabi}, must be smaller than the gravitational force to avoid an accumulation of the atoms at the top of the ellipse, where the rf coupling vanishes~\cite{Morizot2007}.
In the limit $\varepsilon^2\ll 1$, these two conditions can be written approximately\footnote{The exact conditions to fulfil are $\varepsilon < 1$ and $\frac{\Omega_0}{\omega_{\rm rf}}<2\varepsilon\sqrt{1-\varepsilon^2}/(1-3\varepsilon^2)$.} as:
\begin{equation}
m_F\hbar\Omega_0 < Mgr_0 < 2 m_F \hbar \omega_{\rm rf}.
\label{eq:conditions}
\end{equation}

These conditions are always satisfied in our experiments, $\varepsilon$ being always smaller than $0.3$.

Taking into account the small shift due to gravity, the trap centre is slightly offset from the bottom\footnote{The position of the minimum is very close to the isomagnetic surface $r_0$ for large magnetic gradients (small values of $\varepsilon$). More precisely, using the condition on the Rabi coupling, $R$ can be bound as follows: $1<\frac{2R}{r_0}<(1-\varepsilon^2)/(1-3\varepsilon^2)$.} of the ellipsoid~\eqref{eq:ellipsoid}; it lies at the position $(x=0,y=0,z=-R)$ where
\begin{equation}
R = \frac{r_0}{2}\left( 1 + \frac{\varepsilon}{\sqrt{1-\varepsilon^2}}\frac{\Omega_0}{\omega_{\rm rf}} \right).
\label{eq:eqpos}
\end{equation}

For the choice of circular polarization of the rf dressing field, the trap is cylindrically symmetric, and the oscillation frequencies in the horizontal and vertical directions, $\omega_{r}$ and $\omega_z$, obtained from a second order expansion of the potential, are:
\begin{subequations}
  \begin{eqnarray}
 \omega_{r} &=& \sqrt{\frac{g}{4R}}\left[ 1 - \frac{m_F\hbar\Omega_0}{2Mg R}\sqrt{1-\varepsilon^2} \right]^{1/2} ,\\
   \omega_z &=& 2\alpha \sqrt{\frac{m_F\hbar}{M\Omega_0}}\left(1-\varepsilon^2\right)^{3/4}.
  \end{eqnarray}
  \label{eq:freq}
\end{subequations}

The horizontal frequency is essentially the pendulum frequency corresponding to the radius of curvature $4R$ of the isomagnetic surface at the equilibrium position $z=-R$, while the vertical frequency is directly related to the avoided crossing of the adiabatic potential.
The ratio between the oscillation frequencies, to leading order in the small parameters $\varepsilon^2$ and $\varepsilon\eta$, is
\begin{equation}
\frac{\omega_{r}}{\omega_z} \sim \frac{1}{2}\sqrt{\varepsilon\eta[1-\eta/(2\varepsilon)]} < \frac{1}{2}\sqrt{\varepsilon\eta}
\label{eq:ratio}
\end{equation}
where $\eta=\Omega_0/\omega_{\rm rf}<2\varepsilon$, see Eq.~\eqref{eq:conditions}, such that $\omega_{r}/\omega_z$ has to be smaller than $\varepsilon/\sqrt{2}$.
The adiabatic potential described in this paper thus leads to an oblate trap.
For our experimental parameters, the two frequencies are very different, their ratio ranging from 0.1 to 0.01.

Taking advantage of this natural anisotropy, we now discuss the ability of reaching the quasi-two-dimensional regime for a quantum gas confined in the trap. 
As mentioned above, the gas can be considered as quasi two-dimensional if the typical energies involved in the system, the chemical potential $\mu$ and the thermal energy $k_B T$, are less than the transverse, large, trapping energy $\hbar \omega_z$.

To check if the chemical potential is below the transverse frequency, we can estimate it for a three-dimensional condensate confined in a three-dimensional harmonic trap in the mean-field Thomas-Fermi regime. If this value $\mu_{3D}$ is below $\hbar \omega_z$, the hypothesis of a Thomas-Fermi profile along $z$ fails and the gas enters the quasi two-dimensional regime.

The expression for $\mu_{3D}$ is~\cite{Dalfovo1999}
\begin{equation}
\mu_{3D} = \frac{\hbar\bar{\omega}}{2}\left(\frac{15Na}{a_0}\right)^{2/5} \propto \omega_{r}^{4/5}\omega_z^{2/5}
\end{equation} 
where $N$ is the condensate atom number, $a$ is the scattering length, $\bar{\omega}=(\omega_z\omega_{r}^2)^{1/3}$ and $a_0=\sqrt{\hbar/(M\bar{\omega})}$.
$\mu_{3D}/(\hbar\omega_z)$ scales as $\omega_{r}^{4/5}\omega_z^{-3/5}$, and reaching the two-dimensional regime thus requires to lower the frequency ratio $\frac{\omega_{r}}{\omega_z}$.

As $\mu_{3D}$ becomes lower than $\hbar\omega_z$, the true chemical potential $\mu$ can instead be written as the sum of the vertical ground state energy and a quasi-two-dimensional chemical potential, $\mu = \mu_{2D}+\hbar\omega_z/2$. In the spirit of \cite{Dalfovo1999}, $\mu_{2D}$ can be calculated in the Thomas-Fermi regime for the two-dimensional gas from the knowledge of the oscillation frequencies and the number of atoms in the coherent peak. The expression for $\mu_{2D}$ is
\begin{equation}
\mu_{2D} = 2\hbar\omega_{r}\left(\frac{Na}{\sqrt{2\pi}a_z}\right)^{1/2}
\label{eq:mu2D}
\end{equation}
where $a_z=\sqrt{\hbar/(M\omega_z)}$.
Again, this quantity should be small as compared to $\hbar \omega_z$ to ensure the two-dimensional regime for the interacting quantum degenerate gas. It can be shown that $\mu_{2D}/(\hbar\omega_z)\sim[\mu_{3D}/(\hbar\omega_z)]^{5/4}$ such that the two requirements are equivalent~\cite{Hechenblaikner2005}.

Using Eq.~\eqref{eq:ratio}, we can see that a good way to decrease the frequency ratio $\frac{\omega_{r}}{\omega_z}$ is to simultaneously decrease $\varepsilon$ and $\eta$, keeping $\eta/(2\varepsilon)=Mgr_0/(\hbar\Omega_0)$ constant.
This is achieved by ramping simultaneously the magnetic field gradient and the rf frequency, at constant Rabi coupling, to keep the radius $r_0$ constant.
As a result, $\omega_z$ is increased and the frequency ratio scales as $1/\alpha$. The horizontal frequency $\omega_{r}$ barely changes, and the cloud is simply compressed in the vertical direction.

Large vertical oscillation frequencies can be achieved by increasing the magnetic gradient.
We show in section~\ref{sec:2D} that oscillation frequencies of $\omega_z/(2\pi)=2.4$~kHz and $\omega_{r}/(2\pi)=25$~Hz can be obtained in our setup, which is typical for two-dimensional experiments~\cite{Hadzibabic2006,Clade2009a,Plisson2011,Hung2011,Goerlitz2001,Rychtarik2004}.
With these figures, the criterion $\mu_{3D}<\hbar \omega_z$ requires an atom number in the quantum gas below $10^5$, which is easy to meet while keeping good signal to noise ratio for imaging.
Moreover this vertical oscillation frequency corresponds to a temperature of 115~nK, which can easily be reached by forcing evaporation in the dressed trap with a second, weak, rf field~\cite{Garrido2006,Hofferberth2006,KollengodeEaswaran2010}.
Both the superfluid and the thermal fractions will then be in the quasi-two-dimensional regime, which makes this trap well suited for a study of the BKT physics~\cite{Hadzibabic2006}.
 
\section{Experimental implementation and characterization}
\label{sec:BEC}

We now describe the experimental procedure allowing the preparation of quantum degenerate gases in this very anisotropic trap.
An initial atomic sample consisting of $2\times 10^5$ atoms in the $F=1$, $m=-1$ state of $^{87}$Rb at a temperature of 250~nK with a condensate fraction of 0.6 is first produced in a magnetic quadrupole trap plugged by a blue detuned laser beam~\cite{Dubessy2012a}, see figure~\ref{fig:trap}.
The plug beam (wavelength 532~nm, power 6~W) propagates along the $y$ axis and prevents Majorana spin flips in the quadrupole field by shifting the trapping potential minimum horizontally by about 60~$\mu$m to the side of the quadrupole centre.
The Larmor frequency at the position of the plugged trap centre is 250~kHz, corresponding to a bias magnetic field of 350~mG.
The same quadrupole field of vertical ($z$) axis and radial magnetic gradient $b'=55.4$~G$\cdot$cm$^{-1}$ is used both for the plugged trap and the dressed trap, which greatly simplifies the loading procedure, see figure~\ref{fig:trap}. Further compression is possible and the magnetic field gradient can be increased up to 216~G$\cdot$cm$^{-1}$. Detailed experimental procedures for the production and the detection of quantum gases in our setup are given in~\cite{Dubessy2012a}.

We then switch on an rf field of Rabi coupling $\Omega_0= 2\pi\times 40$~kHz at a frequency $\omega_{\rm rf} = 2\pi\times 175$~kHz, below the resonance frequency in the plugged trap, which transfers the $m=-1$ bare state into the $m_F=+1$ dressed state defined in the previous section.
The circularly polarized rf field is produced by a DDS synthesizer, amplified up to 24~dBm by a standard operational amplifier, and feeding two antennas\footnote{The antennas are made of ten loops of 0.71~mm diameter copper wire. They have a square shape with a 16~mm side, and are located at about 10~mm from the atoms.} of axes $x$ and $y$ with a $\pi/2$ phase shift.
To load the atoms into the dressed trap, we ramp up the dressing rf frequency up to 600~kHz in 75~ms, while the intensity of the plug beam is ramped down to zero in 50~ms.
When the rf frequency crosses the resonance at 250~kHz, the atoms are transferred to the upper dressed state, and follow this state adiabatically as the rf frequency is further increased, see figure~\ref{fig:trap}(a).
They remain trapped to an isomagnetic surface of increasing radius $r_0$.
The centre of mass of the cloud is shifted downwards from $A$ to $B$ due to gravity while the plug beam is ramped down, and the condensate reaches its final position at $(0,0,-R)$ at the end of this procedure, where $R=78~\mu$m. The overall transfer efficiency is about 80\%, leading to a final total atom number of $1.6\times 10^5$. During the whole transfer procedure, we can also add a second, weak, rf field at a frequency  of 680~kHz to limit the trap depth and prevent heating. Changing its value allows to adjust the final cloud temperature and the condensate fraction between zero and almost one.

\begin{figure}[t]%
	\centering
 	\includegraphics[width=\linewidth]{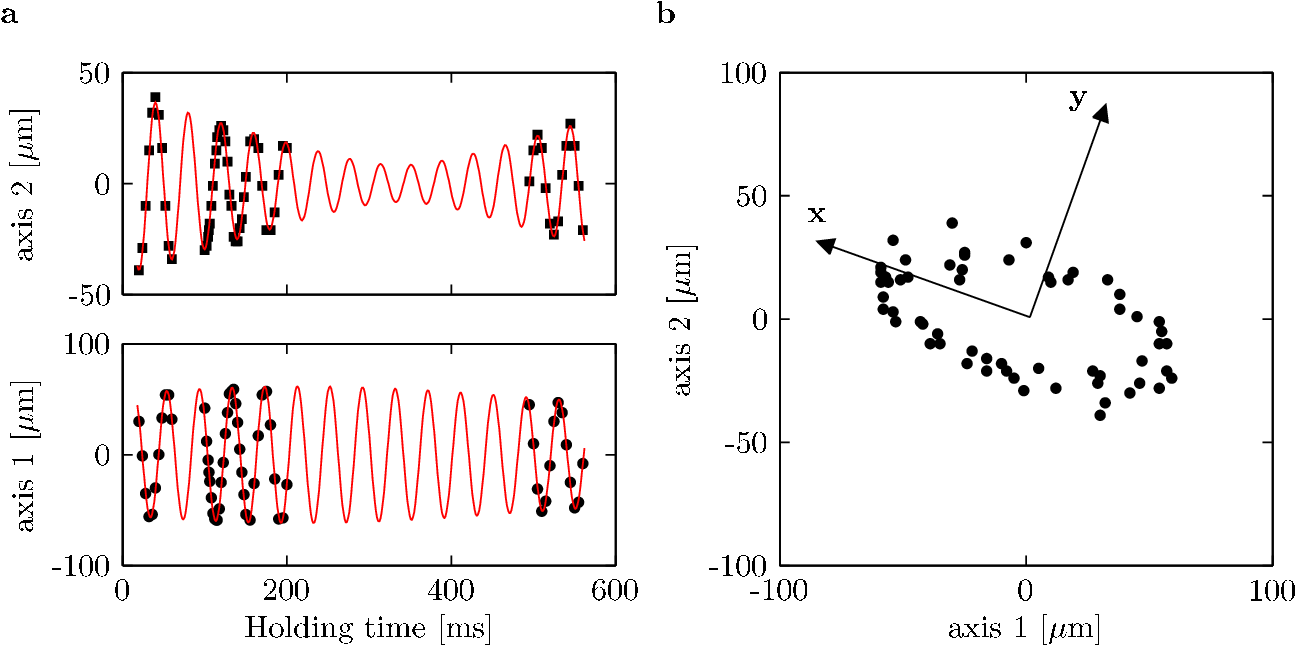}
	\caption{Dipolar oscillation in the dressed quadrupole trap. The experimental parameter are $b'=216$~G$\cdot$cm$^{-1}$, $\omega_{\rm rf} = 2\pi\times 3$~MHz and $\Omega_0= 2\pi\times 21.8$~kHz, corresponding to $R=99.4~\mu$m and a calculated horizontal frequency for an isotropic trap of 24.4~Hz. The axes 1 and 2 correspond to the camera pixel array orientation. (a): The two measured frequencies of $24.7\pm 0.1$~Hz and $25.3\pm 0.1$~Hz are determined by a simultaneous fit of the position of the cloud centre along the $x$ and $y$ axes, with the sum of two sinusoids (solid line). No damping is observable. (b) Trajectory of the centre of mass in the horizontal $xy$ plane.}%
	\label{fig:dipole}
\end{figure}

\begin{center}
\begin{figure}[b]%
	\centering
	\includegraphics{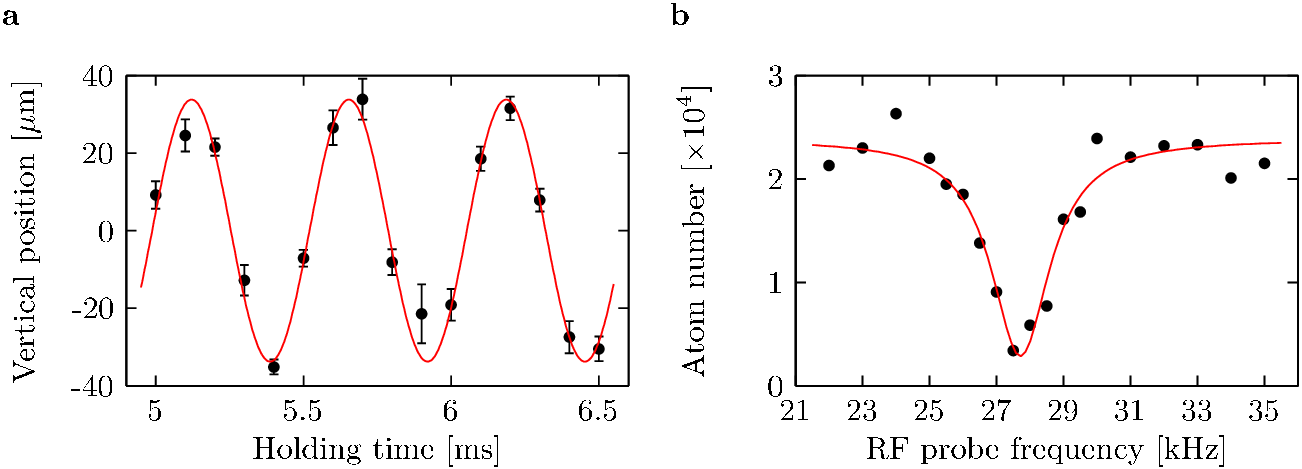}
	\caption{(a): Dipolar oscillation along $z$ in the compressed trap, measured after a 25-ms time-of-flight.
	Each point is the average of three measurements and the error bars estimate the statistical uncertainty.
	The solid line is a sinusoidal fit to the experimental data and gives $1.88\pm0.01$~kHz. Similar data taken with different oscillation amplitudes yield an oscillation frequency of $\omega_z=2\pi\times 1.93\pm0.01$~kHz, corrected for the oscillation amplitude.
	(b) rf spectroscopy in the dressed quadrupole trap in the same conditions.
	The solid line is a Lorentzian fit to the data.
	From this measurement we deduce a Rabi coupling of $27.7\pm 0.1$~kHz.}
	\label{fig:dressfreqz}
\end{figure}
\end{center}

In our experiments in the dressed quadrupole trap, we always start from this configuration, with a moderate magnetic gradient, where the quantum gas is anisotropic but still three-dimensional, see figure~\ref{fig:trap}(c). We have performed a series of measurements to determine the trap characteristics with these figures.
The oscillation frequencies depend on the magnetic gradient, the rf frequency and the rf coupling, see Eq.~\eqref{eq:freq}.
The magnetic gradient was measured previously~\cite{Dubessy2012a}.
While the rf frequency is very well known, the Rabi coupling experienced by the atoms can be measured \textit{in situ} with a resolution of 0.5~kHz by rf spectroscopy~\cite{KollengodeEaswaran2010,Hofferberth2007}.
The number of atoms remaining in the dressed trap is recorded as a function of the frequency of a second, weak, rf field, which induces resonant spin flips to an un-trapped dressed state when its frequency is equal to the Rabi coupling $\Omega_0/(2\pi)$ at the position of the atoms, see for example figure~\ref{fig:dressfreqz}(b).
In our setup the Rabi coupling can be adjusted between 5 and 50~kHz through the input rf power.

For the two-dimensional experiments, it is important to measure the vertical oscillation frequency.
This is done by inducing a sudden vertical displacement of the trap centre and recording the dipolar oscillation in the position of the cloud along the $z$ axis after a 25~ms time-of-flight. 
The vertical displacement is obtained by increasing the value of the rf frequency by 10~kHz in 50~$\mu$s, a time shorter than the oscillation period but still adiabatic with respect to the spin variables.
With an rf frequency of 600~kHz and Rabi coupling 43~kHz, and a gradient of 55.4~G$\cdot$cm$^{-1}$, we measure $\omega_z/(2\pi)=400\pm$10~Hz, in good agreement with the value of 380~Hz deduced from Eq.~\eqref{eq:freq}.
The horizontal oscillation frequencies are deduced from a resonant excitation of the dipolar mode.
We obtain  two peaks at $26.8\pm0.2$~Hz and $27.5\pm0.2$~Hz, corresponding to a small 2.5\% in-plane anisotropy, with an average value in very good agreement with the prediction 27.1~Hz of Eq.~\eqref{eq:freq}. Given the typical condensed atom number between $4\times 10^4$ and $8\times 10^4$, this corresponds to the case where we have a three-dimensional Bose-Einstein condensate, with $\mu_{3D}/(\hbar\omega_z)$ ranging from 3.7 to 4.9.
\onecolumngrid

\twocolumngrid

The horizontal oscillation frequencies can also be determined by exciting the dipolar oscillation and recording directly the in-plane cloud position as a function of time, as demonstrated in figure~\ref{fig:dipole} for another set of trap parameters. This measurement confirms the very small anisotropy. Moreover, the dipolar oscillation can be monitored for half a second without detecting any damping~\cite{vanEs2008}. This feature reveals the excellent harmonic character of the dressed quadrupole trap.

Another remarkable feature of this trap is the low heating rate and the long lifetime which can be achieved. We recorded the total number of atoms in the cloud as a function of time.
The lifetime is limited by the background gas collisions to more than 200\,s for a Rabi coupling above 20~kHz.
We also measured the number of atoms in the coherent central peak.
The lifetime of the degenerate gas, recorded in the presence of an rf-knife 200~kHz above the trap bottom, reaches 10\,s and is limited by three-body losses.

In the same experimental conditions, the heating rate is as low as 4~nK$\cdot$s$^{-1}$. It can be attributed to a small amplitude noise~\cite{Morizot2008} of the fields produced by the two independent antennas, responsible for a heating of 2~nK$\cdot$s$^{-1}$ each.

These figures depend on the Rabi coupling between magnetic states. At very small values of the Rabi coupling, Landau-Zener transitions may occur and limit the lifetime. We observe a reduction in the lifetime of thermal atoms for Rabi frequencies below 20~kHz for a magnetic gradient of 55.4~G$\cdot$cm$^{-1}$. We expect that the lifetime will be lower with a larger magnetic gradient. For the whole set of data presented in this paper, the lifetime is always above 20~s. The issue of lifetime in dressed traps has been addressed in part in Refs.~\cite{Zobay2004,Brink2006}. However, quantitative comparisons with experiments, in both the cases of thermal atoms and quantum degenerate gases, would require further theoretical and experimental studies.

\section{A two-dimensional quantum gas in the dressed trap}
\label{sec:2D}
As demonstrated in the previous sections, the dressed quadrupole trap is naturally oblate with a large anisotropy. In the horizontal plane, it is very well described by an harmonic potential. Moreover, it exhibits a very long lifetime and a low heating rate. 
In this section, we show that we are able to compress the gas in the vertical direction up to very large values of the anisotropy and to reach the two-dimensional regime for the trapped quantum gas.

As discussed at the end of section~\ref{sec:trap}, we increase the vertical trapping frequency while keeping the ellipsoid radius fixed at $R=78~\mu$m. The rf frequency is ramped from $600$~kHz to $2.336$~MHz in 300~ms, and simultaneously the magnetic gradient $b'$ is ramped proportionally from 55.4 to 216~G$\cdot$cm$^{-1}$. The rf-coupling $\Omega_0$ slightly changes during this procedure, due to the frequency dependence of the amplifier gain.
A spectroscopic measurement in the compressed trap at 2.336~MHz gives a smaller value of $27.7\pm 0.1$~kHz, which contributes to increasing the vertical frequency further, see figure~\ref{fig:dressfreqz}(b).
In these conditions, we reach a vertical frequency of $1.93\pm0.01$~kHz, measured by dipolar excitation, and corrected from the oscillation amplitude, see figure~\ref{fig:dressfreqz}(a).
The corresponding horizontal frequency is $\omega_{r}=2\pi\times 26$~Hz.

\begin{center}
\begin{figure}[t]%
	\centering
 	\includegraphics{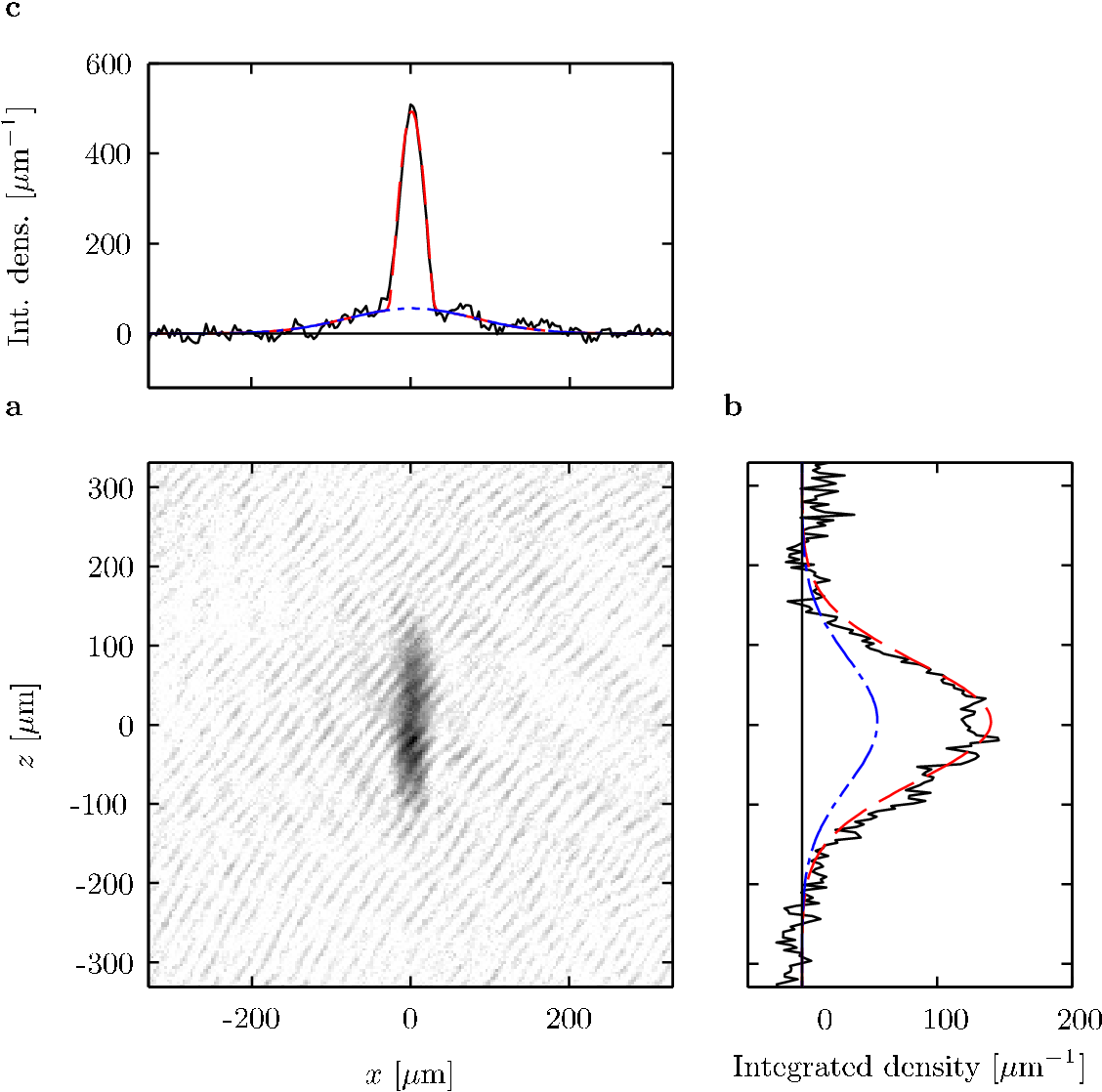}%
	\caption{
	(a) Absorption imaging of the atomic cloud released from the strongly confined dressed quadrupole trap, after a 25-ms time-of-flight. The trap frequencies are $\omega_{r}=2\pi\times 26$~Hz and $\omega_z=2\pi\times$~1.93~kHz. The density profile is fitted by the product of a Gaussian with an integrated Thomas-Fermi profile, the cloud orientation being a free fit parameter. A background thermal fraction is also taken into account with another Gaussian distribution. (b) and (c): Integrated density profiles and integrated two-dimensional fits along the $x$ and $z$ directions, respectively. The horizontal profile of the coherent fraction is an integrate parabola, and originates from the initial two-dimensional parabolic Thomas-Fermi profile in the trap. The corresponding integrated $z$-profile is Gaussian, the atoms being initially confined to the vertical ground state of the anisotropic trap.}%
	\label{fig:tof}
\end{figure}
\end{center}

An adiabatic compression of the cloud results in an increase of the temperature.
We limit this increase by applying a weak rf knife 64~kHz above the dressing frequency during the whole compression phase, leading to evaporative cooling of the sample.
In this way, we prepare almost purely degenerate gases of typically $2\times 10^4$ atoms, see figure~\ref{fig:tof}.
We computed the two-dimensional chemical potential of the gas using Eq.~\eqref{eq:mu2D}. By changing the atom number and the rf frequency at the maximum gradient, we were able to adjust the ratio $\mu_{2D}/(\hbar \omega_z)$ between 0.44 and 0.2, well into the two-dimensional regime.

Figure~\ref{fig:tof}(a) presents an absorption image of the atomic cloud released from the compressed trap, after a 25~ms time of flight.
From this picture we can infer the total atom number, the temperature, the coherent fraction and the density profiles.
The cloud after expansion exhibits a bimodal distribution. The coherent fraction is extremely anisotropic: its density profile has a Thomas-Fermi parabolic shape in the horizontal directions whereas it exhibits a Gaussian shape in the vertical direction, which is a consequence of the confinement to the ground state in the vertical direction. The thermal gas is marginally two-dimensional for this data set, the temperature being $T=112$~nK such that $k_B T/(\hbar\omega_z)=1.2$. Further evaporative cooling leads to colder samples, with no discernible thermal fraction. For these samples, the temperature is estimated by extrapolating the measured temperature as a function of trap depth.

\begin{figure}[t]%
	\centering
 	\includegraphics{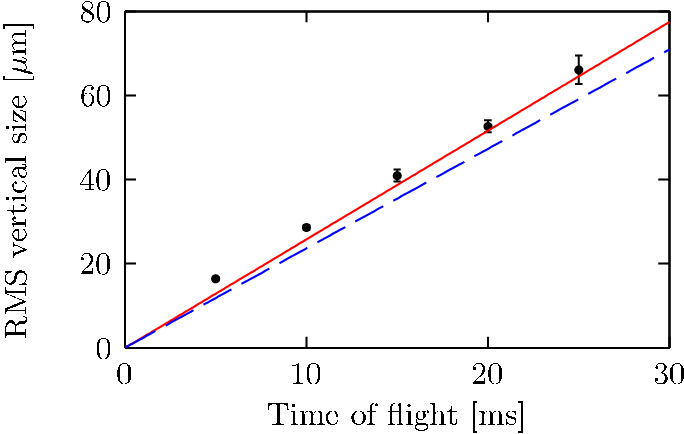}%
	\caption{Vertical rms size of the expanding quasi two-dimensional gas released from the strongly anisotropic trap. Each experimental point (dots) gives the rms size averaged over three shots, with the statistical uncertainty indicated by the error bars. The points are compared with the expansion of the ground state of the harmonic oscillator (dashed blue line) and with the scaling model of Ref.~\cite{Hechenblaikner2005} (solid red line).}%
	\label{fig:expansion}
\end{figure}

In this regime the expansion of the quasi two-dimensional degenerate Bose gas can be analytically described using a scaling ansatz~\cite{Hechenblaikner2005}. We record the time-of-flight expansion of a quasi pure two-dimensional quantum gas with $N=10^4$ atoms at a temperature of 126~nK, see figure~\ref{fig:expansion}. For this data set, the vertical oscillation frequency has been increased up to 2.4~kHz by reducing the Rabi coupling to 18.3~kHz, such that $k_BT\simeq \hbar\omega_z$. The rf dressing frequency is 3~MHz and the horizontal oscillation frequency is 24.6~Hz, corresponding to $\mu_{2D}/(\hbar \omega_z)=0.2$. We compute the rms width of the density profile along $z$ and compare it both to the expansion of an ideal gas in the ground state of the vertical harmonic oscillator and to the rms density width given by a Gaussian ansatz expansion~\cite{Hechenblaikner2005}. Our data agree well with the model taking into account the interactions. It deviates significantly only for the short time-of-flight data, for which the large atomic density saturates the absorption in the cloud centre and results in a systematic overestimation of the rms width.

\onecolumngrid
\twocolumngrid

\section{Two-dimensional collective modes}
\label{sec:modes}

In this last part, we demonstrate a very natural application of the dressed quadrupole trap. Its smoothness and its harmonic character make this trap very well suited for the study of the collective modes of a two-dimensional trapped gas.
Collective modes are an excellent tool to probe the effect of interactions in a quantum gas~\cite{PitaevskiiStringari}, including superfluidity~\cite{Marago2000}, the measurement of the equation of state~\cite{Vogt2012} or dipolar interactions~\cite{Bismut2010}. In particular, the monopole, or breathing, mode is sensitive to the gas density and this allows to deduce the equation of state $\mu(n)$~\cite{Vogt2012}. On the other hand, the quadrupole mode is a signature of the hydrodynamic regime for the quantum gas~\cite{PitaevskiiStringari}. In the following, we show that through controlled deformations of the trap potential we are able to excite the low lying collective modes of the degenerate gas and measure their frequency and quality factor.

We first present a measurement of the monopole mode at $\omega_M = 2\omega_{r}$.
Measuring the monopole mode is of particular interest for a two-dimensional gas.
Due to a scaling invariance~\cite{Hung2011}, this mode is expected to be independent of the amplitude and present an absence of damping~\cite{Pitaevskii1997,Chevy2002}.
In addition, it has been suggested to detect a quantum anomaly~\cite{Olshanii2010} in two dimensions which is responsible for a very small frequency shift.

\begin{figure}[b]%
	\centering
 	\includegraphics{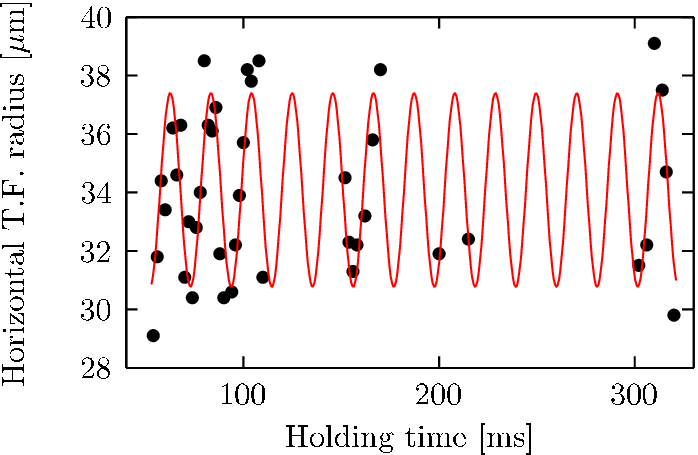}
	\caption{Observation of the monopole mode at $2\omega_{r}$ in the two-dimensional Bose gas.
	The solid line is a sinusoidal fit to the data.
	The measured frequency is $48.1\pm 0.2$~Hz, for a trapping frequency of $24.2\pm0.1$~Hz, measured simultaneously.
	No damping is observable on this time scale.
	}
	\label{fig:monopole}
\end{figure}

The excitation of the monopole mode proceeds as follows. First, starting from the compressed trapped gas described above, we expand the rf-dressed bubble radius by increasing the rf frequency from 2.336 up to 3.336~MHz within 5~ms.
The vertical displacement of the trap centre induced by this frequency sweep is slow compared to the vertical confinement frequency and the atoms follow adiabatically.
However the subsequent change in $\omega_r$ is not adiabatic with respect to the in-plane dynamics: the gas is then placed in an out-of-equilibrium state where its radius is smaller than the equilibrium one.
Since during this procedure the trap cylindrical symmetry is preserved, this results mainly in an excitation of the monopole mode, see figure~\ref{fig:monopole}.
The oscillation frequency $\omega_{r}$ is measured simultaneously from a small residual centre of mass oscillation.  We measure a monopole frequency of $48.1\pm0.2$~Hz for the oscillation of the cloud radius, which is very close to the expected $2\omega_r/(2\pi)= 48.4\pm0.2$~Hz predicted by the theory~\cite{Pitaevskii1997,Stringari1998a}.
The monopole oscillation does not show any sign of damping over half a second. This shows that the quality factor $Q = \omega_M/\Gamma_M$, ratio of the monopole angular frequency to the  damping rate, exceeds 150.
The current experimental resolution doesn't allow for a search for the quantum anomaly, which is at the level of $2\times 10^{-3}$ for our trap parameters~\cite{Olshanii2010}. Furthermore, the quantum anomaly shift will be masked, at the $10^{-2}$ level, by beyond two-dimensional effects originating from a finite value of the ratio $\mu_{2D}/(\hbar\omega_z)$~\cite{Xia2012,Olshanii2012}.

We now turn to the study of the quadrupole mode.
This mode consists in out-of-phase oscillations of the cloud radii along two orthogonal directions.
In our experiment, the thermal gas is always in the collisionless regime. This implies that, while the breathing mode also exists at a frequency $2\omega_{r}$ for the thermal gas, the quadrupole mode can only be observed when a superfluid fraction is present.
In order to excite this mode, we must break the cylindrical symmetry of the trap and induce an anisotropy in the trap potential.
This is done by switching the rf-dressing to another configuration where both antennas are driven in phase to produce the dressing field.

\begin{figure}[t]%
	\centering
 	\includegraphics[width=\linewidth]{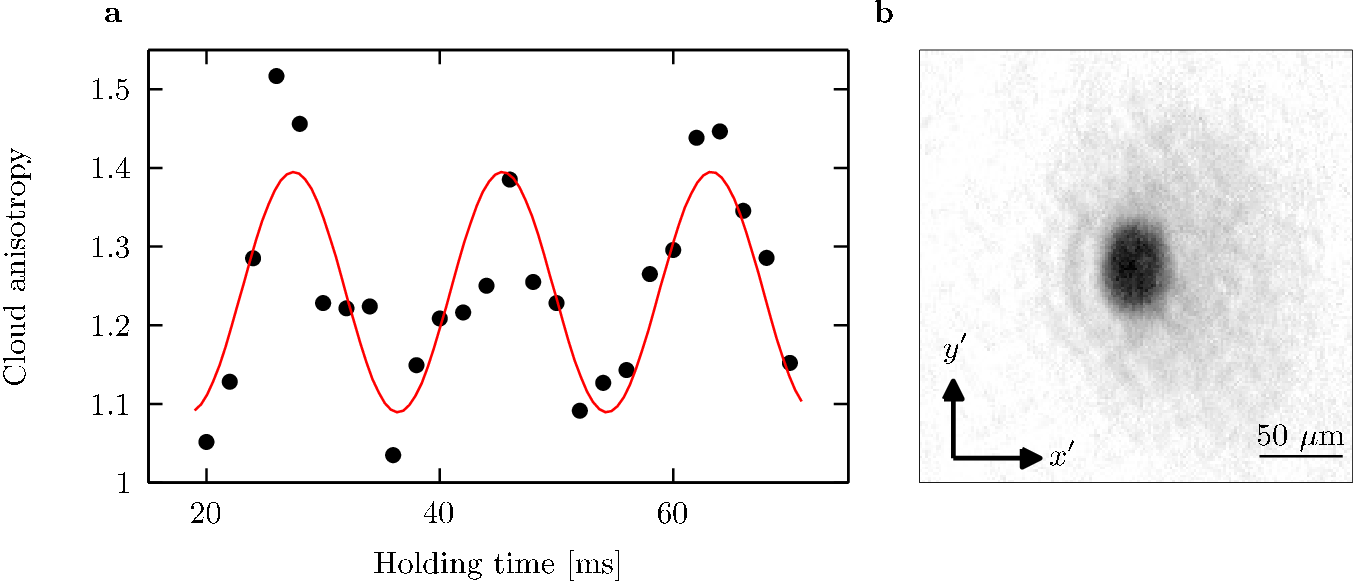}
	\caption{
	(a): Observation of the quadrupole mode in an anisotropic trap: the coherent central part of the cloud density profile exhibits oscillation of its anisotropy at a single frequency when properly excited.
	The solid line is a sinusoidal fit to the data.
	The measured frequency is $56\pm1$~Hz.
	(b): In situ (1 ms time of flight) imaging of the gas density profile by a home made objective (magnification $\times8$, resolution $\simeq4~\mu$m), with probe propagating along $z$.
	In order to reduce the optical thickness only about 10\% of the atoms are pumped on the cycling transition before the imaging pulses.
	The density profile exhibits a bimodal shape with a coherent central feature well described by a Thomas-Fermi profile on top of a larger Gaussian pedestal.
	}%
	\label{fig:quadrupole}
\end{figure}

In this situation, the rf field is linearly polarized along a direction $y'$ at 45 degrees between $x$ and $y$. The degeneracy of the in-plane trapping frequency is lifted, according to Eq.~\eqref{eq:linear}, see Appendix. The in-plane anisotropy can be increased either by reducing the bubble radius $R$ or by increasing the Rabi coupling. We chose to reduce $R$ by tuning the rf frequency to 1~MHz while keeping the Rabi coupling constant to keep the same vertical oscillation frequency. In these conditions, we are able to load degenerate samples with almost no thermal fraction. The temperature estimated from the trap depth is 50~nK. The clouds exhibit an in-plane anisotropy of 1.2, as revealed by in-situ imaging, see figure~\ref{fig:quadrupole}(b). By monitoring small in-situ dipolar oscillations of the cloud, we measure the two trap frequencies: $\omega_{x'}=2\pi\times 45\pm 1$~Hz and $\omega_{y'}=2\pi\times 37.8\pm 0.5$~Hz, consistent with the cloud radii. In this anisotropic trap, the expected frequency for the quadrupole mode, derived from a Castin-Dum analysis~\cite{Castin1996} adapted to two dimensions, is
\begin{equation}
\omega_Q^2 = \frac{3}{2}\left(\omega_{x'}^2 + \omega_{y'}^2\right) - \sqrt{\omega_{x'}^2 \omega_{y'}^2 + \frac{9}{4}\left(\omega_{x'}^2 - \omega_{y'}^2\right)^2}.
\label{eq:quadmode}
\end{equation}

The excitation of the quadrupole mode in this trap proceeds as follows.
The gas is initially at rest in the symmetric trap with circularly polarized dressing field.
We then rapidly change the relative phase between the two antennas, from $\pi/2$ to $\pi$ in 0.4~ms, thus going to the final situation where the dressing field is linearly polarized and the trap is anisotropic.
The initial values of the cloud radii do not correspond to equilibrium in this new trap configuration, one being larger and the other being smaller, which results in out of phase oscillations of the radii at constant cloud volume.
We observe these oscillations by measuring the cloud in-plane anisotropy as a function of time: this quantity oscillates at the quadrupole mode frequency, as shown in figure~\ref{fig:quadrupole}(a).
We measure a quadrupole mode frequency of $56\pm1$~Hz in good agreement with the prediction of $57\pm 1$~Hz, deduced from Eq.~\eqref{eq:quadmode}.

\section{Conclusion}
\label{sec:conclusion}
In summary, we have demonstrated for the first time the confinement of a quantum gas to two dimensions in a magnetic trap. Magnetic traps are very stable, and are not subject to photon scattering or short scale defects as can occur if scattered light is present. Moreover, the trap frequencies are in excellent agreement with those predicted from the simple knowledge of the magnetic gradient and rf characteristics, making the trap geometry well under experimental control. We have shown that the trap is indeed extremely smooth, sustaining dipole oscillations for about one second.

This trap is very well suited for the study of collective modes of a quantum gas. We present data for the monopole and quadrupole mode, in a regime of low temperature.  For this two-dimensional gas, dependence of the mode frequencies on temperature or on the two-dimensional parameter $\mu_{2D}/(\hbar\omega_z)$ are expected~\cite{Salasnich2004} but, up to date, very few experimental data have been published for the bosonic case~\cite{Xia2012}. A shift in the monopole mode frequency, in particular, is a signature of the equation of state. In turn, the equation of state depends on a transverse mean field shift~\cite{Hechenblaikner2005}, or on subtle quantum effects~\cite{Olshanii2010}. A systematic study of these frequency shifts will be the subject of future work.

\appendix

\section{Anisotropic two-dimensional gas}
We consider here a linearly polarized rf field $\mathbf{B}_{\rm rf}(\mathbf{r},t) = B_1\left[\cos(\omega_{\rm rf}t)\,\mathbf{e}_{x}-\cos(\omega_{\rm rf}t)\,\mathbf{e}_{y}\right] = B'_1\cos(\omega_{\rm rf}t)\,\mathbf{e}_{y'}$ along a direction $y'$ in the $xy$ plane making an angle $\theta$ with $y$, i.e. $\mathbf{e}_{y'} = -\sin\theta\mathbf{e}_x + \cos\theta\mathbf{e}_y$. The orthogonal direction in the plane is labeled $x'$, with $\mathbf{e}_{x'} = \cos\theta\mathbf{e}_x + \sin\theta\mathbf{e}_y$. We now define $\Omega_0$ as the maximum rf amplitude at the bottom of the isomagnetic ellipsoid, $\hbar\Omega_0 = \frac{1}{2}|g_F| \mu_B B'_1$. The effective Rabi coupling depends on \textbf{r} as
\begin{equation}
\Omega(\textbf{r}) = \Omega_0\frac{\sqrt{x'^2+4z^2}}{r_e(\textbf{r})}.
\end{equation}
The dressed quadrupole trap now has three non degenerate frequencies along the axes $x'$, $y'$ and $z$. The oscillation frequencies are deduced from a second order expansion of the trapping potential around its minimum. The result is as follows~\cite{Morizot2007}:
\begin{subequations}
  \begin{eqnarray}
 \omega_{x'} &=& \sqrt{\frac{g}{4R}} ,\\
 \omega_{y'} &=& \sqrt{\frac{g}{4R}}\left[ 1 - \frac{m_F\hbar\Omega_0}{Mg R}\sqrt{1-\varepsilon^2} \right]^{1/2} ,\\
   \omega_z &=& 2\alpha \sqrt{\frac{m_F\hbar}{M\Omega_0}}\left(1-\varepsilon^2\right)^{3/4},
  \end{eqnarray}
  \label{eq:linear}
\end{subequations}
with the same equilibrium position $R$, given at Eq.~\eqref{eq:eqpos}. The oscillation frequency in the $x'$ axis depends on $\Omega_0$ only through $R$, because the Rabi coupling is uniform in the whole $x'z$ plane where the static magnetic field is orthogonal to the rf polarization everywhere.

\begin{acknowledgments}
We acknowledge Institut Francilien de Recherche sur les Atomes Froids (IFRAF) for support. We thank Paolo Pedri and Maxim Olshanii for helpful discussions and Fabrice Wiotte for technical assistance with rf sources. LPL is UMR 7538 of CNRS and Paris 13 University.
\end{acknowledgments}

%\bibliographystyle{iopart-num}
%\bibliography{biblioutf8}

\begin{thebibliography}{0}%
\makeatletter
\providecommand \@ifxundefined [1]{%
 \@ifx{#1\undefined}
}%
\providecommand \@ifnum [1]{%
 \ifnum #1\expandafter \@firstoftwo
 \else \expandafter \@secondoftwo
 \fi
}%
\providecommand \@ifx [1]{%
 \ifx #1\expandafter \@firstoftwo
 \else \expandafter \@secondoftwo
 \fi
}%
\providecommand \natexlab [1]{#1}%
\providecommand \enquote  [1]{``#1''}%
\providecommand \bibnamefont  [1]{#1}%
\providecommand \bibfnamefont [1]{#1}%
\providecommand \citenamefont [1]{#1}%
\providecommand \href@noop [0]{\@secondoftwo}%
\providecommand \href [0]{\begingroup \@sanitize@url \@href}%
\providecommand \@href[1]{\@@startlink{#1}\@@href}%
\providecommand \@@href[1]{\endgroup#1\@@endlink}%
\providecommand \@sanitize@url [0]{\catcode `\\12\catcode `\$12\catcode
  `\&12\catcode `\#12\catcode `\^12\catcode `\_12\catcode `\%12\relax}%
\providecommand \@@startlink[1]{}%
\providecommand \@@endlink[0]{}%
\providecommand \url  [0]{\begingroup\@sanitize@url \@url }%
\providecommand \@url [1]{\endgroup\@href {#1}{\urlprefix }}%
\providecommand \urlprefix  [0]{URL }%
\providecommand \Eprint [0]{\href }%
\providecommand \doibase [0]{http://dx.doi.org/}%
\providecommand \selectlanguage [0]{\@gobble}%
\providecommand \bibinfo  [0]{\@secondoftwo}%
\providecommand \bibfield  [0]{\@secondoftwo}%
\providecommand \translation [1]{[#1]}%
\providecommand \BibitemOpen [0]{}%
\providecommand \bibitemStop [0]{}%
\providecommand \bibitemNoStop [0]{.\EOS\space}%
\providecommand \EOS [0]{\spacefactor3000\relax}%
\providecommand \BibitemShut  [1]{\csname bibitem#1\endcsname}%
\let\auto@bib@innerbib\@empty
%</preamble>
\end{thebibliography}%


\begin{thebibliography}{50}%
\makeatletter
\providecommand \@ifxundefined [1]{%
 \@ifx{#1\undefined}
}%
\providecommand \@ifnum [1]{%
 \ifnum #1\expandafter \@firstoftwo
 \else \expandafter \@secondoftwo
 \fi
}%
\providecommand \@ifx [1]{%
 \ifx #1\expandafter \@firstoftwo
 \else \expandafter \@secondoftwo
 \fi
}%
\providecommand \natexlab [1]{#1}%
\providecommand \enquote  [1]{``#1''}%
\providecommand \bibnamefont  [1]{#1}%
\providecommand \bibfnamefont [1]{#1}%
\providecommand \citenamefont [1]{#1}%
\providecommand \href@noop [0]{\@secondoftwo}%
\providecommand \href [0]{\begingroup \@sanitize@url \@href}%
\providecommand \@href[1]{\@@startlink{#1}\@@href}%
\providecommand \@@href[1]{\endgroup#1\@@endlink}%
\providecommand \@sanitize@url [0]{\catcode `\\12\catcode `\$12\catcode
  `\&12\catcode `\#12\catcode `\^12\catcode `\_12\catcode `\%12\relax}%
\providecommand \@@startlink[1]{}%
\providecommand \@@endlink[0]{}%
\providecommand \url  [0]{\begingroup\@sanitize@url \@url }%
\providecommand \@url [1]{\endgroup\@href {#1}{\urlprefix }}%
\providecommand \urlprefix  [0]{URL }%
\providecommand \Eprint [0]{\href }%
\providecommand \doibase [0]{http://dx.doi.org/}%
\providecommand \selectlanguage [0]{\@gobble}%
\providecommand \bibinfo  [0]{\@secondoftwo}%
\providecommand \bibfield  [0]{\@secondoftwo}%
\providecommand \translation [1]{[#1]}%
\providecommand \BibitemOpen [0]{}%
\providecommand \bibitemStop [0]{}%
\providecommand \bibitemNoStop [0]{.\EOS\space}%
\providecommand \EOS [0]{\spacefactor3000\relax}%
\providecommand \BibitemShut  [1]{\csname bibitem#1\endcsname}%
\let\auto@bib@innerbib\@empty
%</preamble>
\bibitem [{\citenamefont {Pricoupenko}\ \emph {et~al.}(2004)\citenamefont
  {Pricoupenko}, \citenamefont {Perrin},\ and\ \citenamefont
  {Olshanii}}]{Houches2003}%
  \BibitemOpen
  \bibinfo {editor} {\bibfnamefont {L.}~\bibnamefont {Pricoupenko}}, \bibinfo
  {editor} {\bibfnamefont {H.}~\bibnamefont {Perrin}}, \ and\ \bibinfo {editor}
  {\bibfnamefont {M.}~\bibnamefont {Olshanii}},\ eds.,\ \href
  {http://jp4.journaldephysique.org/index.php?option=com_toc&url=/articles/jp4/abs/2004/04/contents/contents.html}
  {\emph {\bibinfo {title} {Proceedings of the {E}uroschool on quantum gases in
  low dimensions, Les Houches 2003}}},\ Vol.\ \bibinfo {volume} {116}\
  (\bibinfo  {publisher} {J. Phys. IV},\ \bibinfo {year} {2004})\BibitemShut
  {NoStop}%
\bibitem [{\citenamefont {Bloch}\ \emph {et~al.}(2008)\citenamefont {Bloch},
  \citenamefont {Dalibard},\ and\ \citenamefont {Zwerger}}]{Bloch2008RMP}%
  \BibitemOpen
  \bibfield  {author} {\bibinfo {author} {\bibfnamefont {I.}~\bibnamefont
  {Bloch}}, \bibinfo {author} {\bibfnamefont {J.}~\bibnamefont {Dalibard}}, \
  and\ \bibinfo {author} {\bibfnamefont {W.}~\bibnamefont {Zwerger}},\ }\href
  {\doibase 10.1103/RevModPhys.80.885} {\bibfield  {journal} {\bibinfo
  {journal} {Rev. Mod. Phys.}\ }\textbf {\bibinfo {volume} {80}},\ \bibinfo
  {pages} {885} (\bibinfo {year} {2008})}\BibitemShut {NoStop}%
\bibitem [{\citenamefont {Tonks}(1936)}]{Tonks1936}%
  \BibitemOpen
  \bibfield  {author} {\bibinfo {author} {\bibfnamefont {L.}~\bibnamefont
  {Tonks}},\ }\href {\doibase 10.1103/PhysRev.50.955} {\bibfield  {journal}
  {\bibinfo  {journal} {Phys. Rev.}\ }\textbf {\bibinfo {volume} {50}},\
  \bibinfo {pages} {955} (\bibinfo {year} {1936})}\BibitemShut {NoStop}%
\bibitem [{\citenamefont {Girardeau}(1960)}]{Girardeau1960}%
  \BibitemOpen
  \bibfield  {author} {\bibinfo {author} {\bibfnamefont {M.}~\bibnamefont
  {Girardeau}},\ }\href {\doibase 10.1063/1.1703687} {\bibfield  {journal}
  {\bibinfo  {journal} {J. Math. Phys.}\ }\textbf {\bibinfo {volume} {1}},\
  \bibinfo {pages} {516} (\bibinfo {year} {1960})}\BibitemShut {NoStop}%
\bibitem [{\citenamefont {Berezinskii}(1972)}]{Berezinskii1972}%
  \BibitemOpen
  \bibfield  {author} {\bibinfo {author} {\bibfnamefont {V.}~\bibnamefont
  {Berezinskii}},\ }\href
  {http://www.jetp.ac.ru/cgi-bin/index/e/34/3/p610?a=list} {\bibfield
  {journal} {\bibinfo  {journal} {Sov. Phys. JETP-USSR}\ }\textbf {\bibinfo
  {volume} {34}},\ \bibinfo {pages} {610} (\bibinfo {year} {1972})}\BibitemShut
  {NoStop}%
\bibitem [{\citenamefont {Kosterlitz}\ and\ \citenamefont
  {Thouless}(1973)}]{Kosterlitz1973}%
  \BibitemOpen
  \bibfield  {author} {\bibinfo {author} {\bibfnamefont {J.}~\bibnamefont
  {Kosterlitz}}\ and\ \bibinfo {author} {\bibfnamefont {D.}~\bibnamefont
  {Thouless}},\ }\href {http://stacks.iop.org/0022-3719/6/i=7/a=010} {\bibfield
   {journal} {\bibinfo  {journal} {J. Phys. C : Solid State Phys.}\ }\textbf
  {\bibinfo {volume} {6}},\ \bibinfo {pages} {1181} (\bibinfo {year}
  {1973})}\BibitemShut {NoStop}%
\bibitem [{\citenamefont {Prokof'ev}\ \emph {et~al.}(2001)\citenamefont
  {Prokof'ev}, \citenamefont {Ruebenacker},\ and\ \citenamefont
  {Svistunov}}]{Prokofev2001}%
  \BibitemOpen
  \bibfield  {author} {\bibinfo {author} {\bibfnamefont {N.}~\bibnamefont
  {Prokof'ev}}, \bibinfo {author} {\bibfnamefont {O.}~\bibnamefont
  {Ruebenacker}}, \ and\ \bibinfo {author} {\bibfnamefont {B.}~\bibnamefont
  {Svistunov}},\ }\href {\doibase 10.1103/PhysRevLett.87.270402} {\bibfield
  {journal} {\bibinfo  {journal} {Phys. Rev. Lett.}\ }\textbf {\bibinfo
  {volume} {87}},\ \bibinfo {pages} {270402} (\bibinfo {year}
  {2001})}\BibitemShut {NoStop}%
\bibitem [{\citenamefont {Tolra}\ \emph {et~al.}(2004)\citenamefont {Tolra},
  \citenamefont {O'Hara}, \citenamefont {Huckans}, \citenamefont {Phillips},
  \citenamefont {Rolston},\ and\ \citenamefont {Porto}}]{Laburthe2004a}%
  \BibitemOpen
  \bibfield  {author} {\bibinfo {author} {\bibfnamefont {B.~L.}\ \bibnamefont
  {Tolra}}, \bibinfo {author} {\bibfnamefont {K.~M.}\ \bibnamefont {O'Hara}},
  \bibinfo {author} {\bibfnamefont {J.~H.}\ \bibnamefont {Huckans}}, \bibinfo
  {author} {\bibfnamefont {W.~D.}\ \bibnamefont {Phillips}}, \bibinfo {author}
  {\bibfnamefont {S.~L.}\ \bibnamefont {Rolston}}, \ and\ \bibinfo {author}
  {\bibfnamefont {J.~V.}\ \bibnamefont {Porto}},\ }\href {\doibase
  10.1103/PhysRevLett.92.190401} {\bibfield  {journal} {\bibinfo  {journal}
  {Phys. Rev. Lett.}\ }\textbf {\bibinfo {volume} {92}},\ \bibinfo {pages}
  {190401} (\bibinfo {year} {2004})}\BibitemShut {NoStop}%
\bibitem [{\citenamefont {Kinoshita}\ \emph {et~al.}(2004)\citenamefont
  {Kinoshita}, \citenamefont {Wenger},\ and\ \citenamefont
  {Weiss}}]{Kinoshita2004}%
  \BibitemOpen
  \bibfield  {author} {\bibinfo {author} {\bibfnamefont {T.}~\bibnamefont
  {Kinoshita}}, \bibinfo {author} {\bibfnamefont {T.}~\bibnamefont {Wenger}}, \
  and\ \bibinfo {author} {\bibfnamefont {D.~S.}\ \bibnamefont {Weiss}},\ }\href
  {\doibase 10.1126/science.1100700} {\bibfield  {journal} {\bibinfo  {journal}
  {Science}\ }\textbf {\bibinfo {volume} {305}},\ \bibinfo {pages} {1125}
  (\bibinfo {year} {2004})},\ \Eprint
  {http://arxiv.org/abs/http://www.sciencemag.org/content/305/5687/1125.full.pdf}
  {http://www.sciencemag.org/content/305/5687/1125.full.pdf} \BibitemShut
  {NoStop}%
\bibitem [{\citenamefont {Paredes}\ \emph {et~al.}(2004)\citenamefont
  {Paredes}, \citenamefont {Widera}, \citenamefont {Murg}, \citenamefont
  {Mandel}, \citenamefont {Folling}, \citenamefont {Cirac}, \citenamefont
  {Shlyapnikov}, \citenamefont {Hansch},\ and\ \citenamefont
  {Bloch}}]{Paredes2004a}%
  \BibitemOpen
  \bibfield  {author} {\bibinfo {author} {\bibfnamefont {B.}~\bibnamefont
  {Paredes}}, \bibinfo {author} {\bibfnamefont {A.}~\bibnamefont {Widera}},
  \bibinfo {author} {\bibfnamefont {V.}~\bibnamefont {Murg}}, \bibinfo {author}
  {\bibfnamefont {O.}~\bibnamefont {Mandel}}, \bibinfo {author} {\bibfnamefont
  {S.}~\bibnamefont {Folling}}, \bibinfo {author} {\bibfnamefont
  {I.}~\bibnamefont {Cirac}}, \bibinfo {author} {\bibfnamefont {G.~V.}\
  \bibnamefont {Shlyapnikov}}, \bibinfo {author} {\bibfnamefont {T.~W.}\
  \bibnamefont {Hansch}}, \ and\ \bibinfo {author} {\bibfnamefont
  {I.}~\bibnamefont {Bloch}},\ }\href {http://dx.doi.org/10.1038/nature02530}
  {\bibfield  {journal} {\bibinfo  {journal} {Nature}\ }\textbf {\bibinfo
  {volume} {429}},\ \bibinfo {pages} {277} (\bibinfo {year}
  {2004})}\BibitemShut {NoStop}%
\bibitem [{\citenamefont {Hadzibabic}\ \emph {et~al.}(2006)\citenamefont
  {Hadzibabic}, \citenamefont {Kr\"uger}, \citenamefont {Cheneau},
  \citenamefont {Battelier},\ and\ \citenamefont {Dalibard}}]{Hadzibabic2006}%
  \BibitemOpen
  \bibfield  {author} {\bibinfo {author} {\bibfnamefont {Z.}~\bibnamefont
  {Hadzibabic}}, \bibinfo {author} {\bibfnamefont {P.}~\bibnamefont
  {Kr\"uger}}, \bibinfo {author} {\bibfnamefont {M.}~\bibnamefont {Cheneau}},
  \bibinfo {author} {\bibfnamefont {B.}~\bibnamefont {Battelier}}, \ and\
  \bibinfo {author} {\bibfnamefont {J.}~\bibnamefont {Dalibard}},\ }\href
  {http://dx.doi.org/10.1038/nature04851} {\bibfield  {journal} {\bibinfo
  {journal} {Nature}\ }\textbf {\bibinfo {volume} {441}},\ \bibinfo {pages}
  {1118} (\bibinfo {year} {2006})}\BibitemShut {NoStop}%
\bibitem [{\citenamefont {Clad\'e}\ \emph {et~al.}(2009)\citenamefont
  {Clad\'e}, \citenamefont {Ryu}, \citenamefont {Ramanathan}, \citenamefont
  {Helmerson},\ and\ \citenamefont {Phillips}}]{Clade2009a}%
  \BibitemOpen
  \bibfield  {author} {\bibinfo {author} {\bibfnamefont {P.}~\bibnamefont
  {Clad\'e}}, \bibinfo {author} {\bibfnamefont {C.}~\bibnamefont {Ryu}},
  \bibinfo {author} {\bibfnamefont {A.}~\bibnamefont {Ramanathan}}, \bibinfo
  {author} {\bibfnamefont {K.}~\bibnamefont {Helmerson}}, \ and\ \bibinfo
  {author} {\bibfnamefont {W.~D.}\ \bibnamefont {Phillips}},\ }\href {\doibase
  10.1103/PhysRevLett.102.170401} {\bibfield  {journal} {\bibinfo  {journal}
  {Phys. Rev. Lett.}\ }\textbf {\bibinfo {volume} {102}},\ \bibinfo {pages}
  {170401} (\bibinfo {year} {2009})}\BibitemShut {NoStop}%
\bibitem [{\citenamefont {Rath}\ \emph {et~al.}(2010)\citenamefont {Rath},
  \citenamefont {Yefsah}, \citenamefont {G\"unter}, \citenamefont {Cheneau},
  \citenamefont {Desbuquois}, \citenamefont {Holzmann}, \citenamefont
  {Krauth},\ and\ \citenamefont {Dalibard}}]{Rath2010}%
  \BibitemOpen
  \bibfield  {author} {\bibinfo {author} {\bibfnamefont {S.~P.}\ \bibnamefont
  {Rath}}, \bibinfo {author} {\bibfnamefont {T.}~\bibnamefont {Yefsah}},
  \bibinfo {author} {\bibfnamefont {K.~J.}\ \bibnamefont {G\"unter}}, \bibinfo
  {author} {\bibfnamefont {M.}~\bibnamefont {Cheneau}}, \bibinfo {author}
  {\bibfnamefont {R.}~\bibnamefont {Desbuquois}}, \bibinfo {author}
  {\bibfnamefont {M.}~\bibnamefont {Holzmann}}, \bibinfo {author}
  {\bibfnamefont {W.}~\bibnamefont {Krauth}}, \ and\ \bibinfo {author}
  {\bibfnamefont {J.}~\bibnamefont {Dalibard}},\ }\href {\doibase
  10.1103/PhysRevA.82.013609} {\bibfield  {journal} {\bibinfo  {journal} {Phys.
  Rev. A}\ }\textbf {\bibinfo {volume} {82}},\ \bibinfo {pages} {013609}
  (\bibinfo {year} {2010})}\BibitemShut {NoStop}%
\bibitem [{\citenamefont {Petrov}\ \emph {et~al.}(2004)\citenamefont {Petrov},
  \citenamefont {Gangardt},\ and\ \citenamefont {Shlyapnikov}}]{Petrov2004}%
  \BibitemOpen
  \bibfield  {author} {\bibinfo {author} {\bibfnamefont {D.}~\bibnamefont
  {Petrov}}, \bibinfo {author} {\bibfnamefont {D.}~\bibnamefont {Gangardt}}, \
  and\ \bibinfo {author} {\bibfnamefont {G.}~\bibnamefont {Shlyapnikov}},\ }in\
  \href {\doibase 10.1051/jp4:2004116001} {\emph {\bibinfo {booktitle}
  {Proceedings of the {E}uroschool on quantum gases in low dimensions, Les
  Houches 2003}}},\ Vol.\ \bibinfo {volume} {116},\ \bibinfo {editor} {edited
  by\ \bibinfo {editor} {\bibfnamefont {L.}~\bibnamefont {Pricoupenko}},
  \bibinfo {editor} {\bibfnamefont {H.}~\bibnamefont {Perrin}}, \ and\ \bibinfo
  {editor} {\bibfnamefont {M.}~\bibnamefont {Olshanii}}}\ (\bibinfo
  {publisher} {J. Phys. IV},\ \bibinfo {year} {2004})\ p.~\bibinfo {pages}
  {5}\BibitemShut {NoStop}%
\bibitem [{\citenamefont {Plisson}\ \emph {et~al.}(2011)\citenamefont
  {Plisson}, \citenamefont {Allard}, \citenamefont {Holzmann}, \citenamefont
  {Salomon}, \citenamefont {Aspect}, \citenamefont {Bouyer},\ and\
  \citenamefont {Bourdel}}]{Plisson2011}%
  \BibitemOpen
  \bibfield  {author} {\bibinfo {author} {\bibfnamefont {T.}~\bibnamefont
  {Plisson}}, \bibinfo {author} {\bibfnamefont {B.}~\bibnamefont {Allard}},
  \bibinfo {author} {\bibfnamefont {M.}~\bibnamefont {Holzmann}}, \bibinfo
  {author} {\bibfnamefont {G.}~\bibnamefont {Salomon}}, \bibinfo {author}
  {\bibfnamefont {A.}~\bibnamefont {Aspect}}, \bibinfo {author} {\bibfnamefont
  {P.}~\bibnamefont {Bouyer}}, \ and\ \bibinfo {author} {\bibfnamefont
  {T.}~\bibnamefont {Bourdel}},\ }\href {\doibase 10.1103/PhysRevA.84.061606}
  {\bibfield  {journal} {\bibinfo  {journal} {Phys. Rev. A}\ }\textbf {\bibinfo
  {volume} {84}},\ \bibinfo {pages} {061606} (\bibinfo {year}
  {2011})}\BibitemShut {NoStop}%
\bibitem [{\citenamefont {Hung}\ \emph {et~al.}(2011)\citenamefont {Hung},
  \citenamefont {Zhang}, \citenamefont {Gemelke},\ and\ \citenamefont
  {Chin}}]{Hung2011}%
  \BibitemOpen
  \bibfield  {author} {\bibinfo {author} {\bibfnamefont {C.-L.}\ \bibnamefont
  {Hung}}, \bibinfo {author} {\bibfnamefont {X.}~\bibnamefont {Zhang}},
  \bibinfo {author} {\bibfnamefont {N.}~\bibnamefont {Gemelke}}, \ and\
  \bibinfo {author} {\bibfnamefont {C.}~\bibnamefont {Chin}},\ }\href
  {http://dx.doi.org/10.1038/nature09722} {\bibfield  {journal} {\bibinfo
  {journal} {Nature}\ }\textbf {\bibinfo {volume} {470}},\ \bibinfo {pages}
  {236} (\bibinfo {year} {2011})}\BibitemShut {NoStop}%
\bibitem [{\citenamefont {Pitaevskii}\ and\ \citenamefont
  {Rosch}(1997)}]{Pitaevskii1997}%
  \BibitemOpen
  \bibfield  {author} {\bibinfo {author} {\bibfnamefont {L.~P.}\ \bibnamefont
  {Pitaevskii}}\ and\ \bibinfo {author} {\bibfnamefont {A.}~\bibnamefont
  {Rosch}},\ }\href {\doibase 10.1103/PhysRevA.55.R853} {\bibfield  {journal}
  {\bibinfo  {journal} {Phys. Rev. A}\ }\textbf {\bibinfo {volume} {55}},\
  \bibinfo {pages} {R853} (\bibinfo {year} {1997})}\BibitemShut {NoStop}%
\bibitem [{\citenamefont {Chevy}\ \emph {et~al.}(2002)\citenamefont {Chevy},
  \citenamefont {Bretin}, \citenamefont {Rosenbusch}, \citenamefont {Madison},\
  and\ \citenamefont {Dalibard}}]{Chevy2002}%
  \BibitemOpen
  \bibfield  {author} {\bibinfo {author} {\bibfnamefont {F.}~\bibnamefont
  {Chevy}}, \bibinfo {author} {\bibfnamefont {V.}~\bibnamefont {Bretin}},
  \bibinfo {author} {\bibfnamefont {P.}~\bibnamefont {Rosenbusch}}, \bibinfo
  {author} {\bibfnamefont {K.~W.}\ \bibnamefont {Madison}}, \ and\ \bibinfo
  {author} {\bibfnamefont {J.}~\bibnamefont {Dalibard}},\ }\href {\doibase
  10.1103/PhysRevLett.88.250402} {\bibfield  {journal} {\bibinfo  {journal}
  {Phys. Rev. Lett.}\ }\textbf {\bibinfo {volume} {88}},\ \bibinfo {pages}
  {250402} (\bibinfo {year} {2002})}\BibitemShut {NoStop}%
\bibitem [{\citenamefont {Allard}\ \emph {et~al.}(2012)\citenamefont {Allard},
  \citenamefont {Plisson}, \citenamefont {Holzmann}, \citenamefont {Salomon},
  \citenamefont {Aspect}, \citenamefont {Bouyer},\ and\ \citenamefont
  {Bourdel}}]{Allard2012}%
  \BibitemOpen
  \bibfield  {author} {\bibinfo {author} {\bibfnamefont {B.}~\bibnamefont
  {Allard}}, \bibinfo {author} {\bibfnamefont {T.}~\bibnamefont {Plisson}},
  \bibinfo {author} {\bibfnamefont {M.}~\bibnamefont {Holzmann}}, \bibinfo
  {author} {\bibfnamefont {G.}~\bibnamefont {Salomon}}, \bibinfo {author}
  {\bibfnamefont {A.}~\bibnamefont {Aspect}}, \bibinfo {author} {\bibfnamefont
  {P.}~\bibnamefont {Bouyer}}, \ and\ \bibinfo {author} {\bibfnamefont
  {T.}~\bibnamefont {Bourdel}},\ }\href {\doibase 10.1103/PhysRevA.85.033602}
  {\bibfield  {journal} {\bibinfo  {journal} {Phys. Rev. A}\ }\textbf {\bibinfo
  {volume} {85}},\ \bibinfo {pages} {033602} (\bibinfo {year}
  {2012})}\BibitemShut {NoStop}%
\bibitem [{\citenamefont {Beeler}\ \emph {et~al.}(2012)\citenamefont {Beeler},
  \citenamefont {Reed}, \citenamefont {Hong},\ and\ \citenamefont
  {Rolston}}]{Beeler2012}%
  \BibitemOpen
  \bibfield  {author} {\bibinfo {author} {\bibfnamefont {M.~C.}\ \bibnamefont
  {Beeler}}, \bibinfo {author} {\bibfnamefont {M.~E.~W.}\ \bibnamefont {Reed}},
  \bibinfo {author} {\bibfnamefont {T.}~\bibnamefont {Hong}}, \ and\ \bibinfo
  {author} {\bibfnamefont {S.~L.}\ \bibnamefont {Rolston}},\ }\href
  {http://stacks.iop.org/1367-2630/14/i=7/a=073024} {\bibfield  {journal}
  {\bibinfo  {journal} {New Journal of Physics}\ }\textbf {\bibinfo {volume}
  {14}},\ \bibinfo {pages} {073024} (\bibinfo {year} {2012})}\BibitemShut
  {NoStop}%
\bibitem [{\citenamefont {G\"orlitz}\ \emph {et~al.}(2001)\citenamefont
  {G\"orlitz}, \citenamefont {Vogels}, \citenamefont {Leanhardt}, \citenamefont
  {Raman}, \citenamefont {Gustavson}, \citenamefont {Abo-Shaeer}, \citenamefont
  {Chikkatur}, \citenamefont {Gupta}, \citenamefont {Inouye}, \citenamefont
  {Rosenband},\ and\ \citenamefont {Ketterle}}]{Goerlitz2001}%
  \BibitemOpen
  \bibfield  {author} {\bibinfo {author} {\bibfnamefont {A.}~\bibnamefont
  {G\"orlitz}}, \bibinfo {author} {\bibfnamefont {J.~M.}\ \bibnamefont
  {Vogels}}, \bibinfo {author} {\bibfnamefont {A.~E.}\ \bibnamefont
  {Leanhardt}}, \bibinfo {author} {\bibfnamefont {C.}~\bibnamefont {Raman}},
  \bibinfo {author} {\bibfnamefont {T.~L.}\ \bibnamefont {Gustavson}}, \bibinfo
  {author} {\bibfnamefont {J.~R.}\ \bibnamefont {Abo-Shaeer}}, \bibinfo
  {author} {\bibfnamefont {A.~P.}\ \bibnamefont {Chikkatur}}, \bibinfo {author}
  {\bibfnamefont {S.}~\bibnamefont {Gupta}}, \bibinfo {author} {\bibfnamefont
  {S.}~\bibnamefont {Inouye}}, \bibinfo {author} {\bibfnamefont
  {T.}~\bibnamefont {Rosenband}}, \ and\ \bibinfo {author} {\bibfnamefont
  {W.}~\bibnamefont {Ketterle}},\ }\href {\doibase
  10.1103/PhysRevLett.87.130402} {\bibfield  {journal} {\bibinfo  {journal}
  {Phys. Rev. Lett.}\ }\textbf {\bibinfo {volume} {87}},\ \bibinfo {pages}
  {130402} (\bibinfo {year} {2001})}\BibitemShut {NoStop}%
\bibitem [{\citenamefont {Rychtarik}\ \emph {et~al.}(2004)\citenamefont
  {Rychtarik}, \citenamefont {Engeser}, \citenamefont {N\"agerl},\ and\
  \citenamefont {Grimm}}]{Rychtarik2004}%
  \BibitemOpen
  \bibfield  {author} {\bibinfo {author} {\bibfnamefont {D.}~\bibnamefont
  {Rychtarik}}, \bibinfo {author} {\bibfnamefont {B.}~\bibnamefont {Engeser}},
  \bibinfo {author} {\bibfnamefont {H.-C.}\ \bibnamefont {N\"agerl}}, \ and\
  \bibinfo {author} {\bibfnamefont {R.}~\bibnamefont {Grimm}},\ }\href
  {\doibase 10.1103/PhysRevLett.92.173003} {\bibfield  {journal} {\bibinfo
  {journal} {Phys. Rev. Lett.}\ }\textbf {\bibinfo {volume} {92}},\ \bibinfo
  {pages} {173003} (\bibinfo {year} {2004})}\BibitemShut {NoStop}%
\bibitem [{\citenamefont {Grimm}\ \emph {et~al.}(2000)\citenamefont {Grimm},
  \citenamefont {Weidem{\"u}ller},\ and\ \citenamefont
  {Ovchinnikov}}]{Grimm2000}%
  \BibitemOpen
  \bibfield  {author} {\bibinfo {author} {\bibfnamefont {R.}~\bibnamefont
  {Grimm}}, \bibinfo {author} {\bibfnamefont {M.}~\bibnamefont
  {Weidem{\"u}ller}}, \ and\ \bibinfo {author} {\bibfnamefont {Y.}~\bibnamefont
  {Ovchinnikov}},\ }\href
  {http://store.elsevier.com/product.jsp?isbn=9780080561530&pagename=search}
  {\bibfield  {journal} {\bibinfo  {journal} {Adv. At. Mol. Opt. Phys.}\
  }\textbf {\bibinfo {volume} {42}},\ \bibinfo {pages} {95} (\bibinfo {year}
  {2000})}\BibitemShut {NoStop}%
\bibitem [{\citenamefont {Gehm}\ \emph {et~al.}(1998)\citenamefont {Gehm},
  \citenamefont {O'Hara}, \citenamefont {Savard},\ and\ \citenamefont
  {Thomas}}]{Gehm1998}%
  \BibitemOpen
  \bibfield  {author} {\bibinfo {author} {\bibfnamefont {M.~E.}\ \bibnamefont
  {Gehm}}, \bibinfo {author} {\bibfnamefont {K.~M.}\ \bibnamefont {O'Hara}},
  \bibinfo {author} {\bibfnamefont {T.~A.}\ \bibnamefont {Savard}}, \ and\
  \bibinfo {author} {\bibfnamefont {J.~E.}\ \bibnamefont {Thomas}},\ }\href
  {\doibase 10.1103/PhysRevA.58.3914} {\bibfield  {journal} {\bibinfo
  {journal} {Phys. Rev. A}\ }\textbf {\bibinfo {volume} {58}},\ \bibinfo
  {pages} {3914} (\bibinfo {year} {1998})},\ \bibinfo {note} {see also:
  {E}rratum: Phys. Rev. A \textbf{61}, 029902 (2000)}\BibitemShut {NoStop}%
\bibitem [{\citenamefont {Zobay}\ and\ \citenamefont
  {Garraway}(2001)}]{Zobay2001}%
  \BibitemOpen
  \bibfield  {author} {\bibinfo {author} {\bibfnamefont {O.}~\bibnamefont
  {Zobay}}\ and\ \bibinfo {author} {\bibfnamefont {B.~M.}\ \bibnamefont
  {Garraway}},\ }\href {\doibase 10.1103/PhysRevLett.86.1195} {\bibfield
  {journal} {\bibinfo  {journal} {Phys. Rev. Lett.}\ }\textbf {\bibinfo
  {volume} {86}},\ \bibinfo {pages} {1195} (\bibinfo {year}
  {2001})}\BibitemShut {NoStop}%
\bibitem [{\citenamefont {Morizot}\ \emph {et~al.}(2007)\citenamefont
  {Morizot}, \citenamefont {Garrido~Alzar}, \citenamefont {Pottie},
  \citenamefont {Lorent},\ and\ \citenamefont {Perrin}}]{Morizot2007}%
  \BibitemOpen
  \bibfield  {author} {\bibinfo {author} {\bibfnamefont {O.}~\bibnamefont
  {Morizot}}, \bibinfo {author} {\bibfnamefont {C.}~\bibnamefont
  {Garrido~Alzar}}, \bibinfo {author} {\bibfnamefont {P.-E.}\ \bibnamefont
  {Pottie}}, \bibinfo {author} {\bibfnamefont {V.}~\bibnamefont {Lorent}}, \
  and\ \bibinfo {author} {\bibfnamefont {H.}~\bibnamefont {Perrin}},\ }\href
  {http://stacks.iop.org/0953-4075/40/i=20/a=004} {\bibfield  {journal}
  {\bibinfo  {journal} {J. Phys. B: At. Mol. Opt. Phys.}\ }\textbf {\bibinfo
  {volume} {40}},\ \bibinfo {pages} {4013} (\bibinfo {year}
  {2007})}\BibitemShut {NoStop}%
\bibitem [{\citenamefont {Dubessy}\ \emph {et~al.}(2012)\citenamefont
  {Dubessy}, \citenamefont {Merloti}, \citenamefont {Longchambon},
  \citenamefont {Pottie}, \citenamefont {Liennard}, \citenamefont {Perrin},
  \citenamefont {Lorent},\ and\ \citenamefont {Perrin}}]{Dubessy2012a}%
  \BibitemOpen
  \bibfield  {author} {\bibinfo {author} {\bibfnamefont {R.}~\bibnamefont
  {Dubessy}}, \bibinfo {author} {\bibfnamefont {K.}~\bibnamefont {Merloti}},
  \bibinfo {author} {\bibfnamefont {L.}~\bibnamefont {Longchambon}}, \bibinfo
  {author} {\bibfnamefont {P.-E.}\ \bibnamefont {Pottie}}, \bibinfo {author}
  {\bibfnamefont {T.}~\bibnamefont {Liennard}}, \bibinfo {author}
  {\bibfnamefont {A.}~\bibnamefont {Perrin}}, \bibinfo {author} {\bibfnamefont
  {V.}~\bibnamefont {Lorent}}, \ and\ \bibinfo {author} {\bibfnamefont
  {H.}~\bibnamefont {Perrin}},\ }\href {\doibase 10.1103/PhysRevA.85.013643}
  {\bibfield  {journal} {\bibinfo  {journal} {Phys. Rev. A}\ }\textbf {\bibinfo
  {volume} {85}},\ \bibinfo {pages} {013643} (\bibinfo {year}
  {2012})}\BibitemShut {NoStop}%
\bibitem [{Note1()}]{Note1}%
  \BibitemOpen
  \bibinfo {note} {The exact conditions to fulfil are $\varepsilon < 1$ and
  $\protect \frac {\Omega _0}{\omega _{\protect \rm rf}}<2\varepsilon \protect
  \sqrt {1-\varepsilon ^2}/(1-3\varepsilon ^2)$.}\BibitemShut {Stop}%
\bibitem [{Note2()}]{Note2}%
  \BibitemOpen
  \bibinfo {note} {The position of the minimum is very close to the isomagnetic
  surface $r_0$ for large magnetic gradients (small values of $\varepsilon $).
  More precisely, using the condition on the Rabi coupling, $R$ can be bound as
  follows: $1<\protect \frac {2R}{r_0}<(1-\varepsilon ^2)/(1-3\varepsilon
  ^2)$.}\BibitemShut {Stop}%
\bibitem [{\citenamefont {Dalfovo}\ \emph {et~al.}(1999)\citenamefont
  {Dalfovo}, \citenamefont {Giorgini}, \citenamefont {Pitaevskii},\ and\
  \citenamefont {Stringari}}]{Dalfovo1999}%
  \BibitemOpen
  \bibfield  {author} {\bibinfo {author} {\bibfnamefont {F.}~\bibnamefont
  {Dalfovo}}, \bibinfo {author} {\bibfnamefont {S.}~\bibnamefont {Giorgini}},
  \bibinfo {author} {\bibfnamefont {L.~P.}\ \bibnamefont {Pitaevskii}}, \ and\
  \bibinfo {author} {\bibfnamefont {S.}~\bibnamefont {Stringari}},\ }\href
  {\doibase 10.1103/RevModPhys.71.463} {\bibfield  {journal} {\bibinfo
  {journal} {Rev. Mod. Phys.}\ }\textbf {\bibinfo {volume} {71}},\ \bibinfo
  {pages} {463} (\bibinfo {year} {1999})}\BibitemShut {NoStop}%
\bibitem [{\citenamefont {Hechenblaikner}\ \emph {et~al.}(2005)\citenamefont
  {Hechenblaikner}, \citenamefont {Krueger},\ and\ \citenamefont
  {Foot}}]{Hechenblaikner2005}%
  \BibitemOpen
  \bibfield  {author} {\bibinfo {author} {\bibfnamefont {G.}~\bibnamefont
  {Hechenblaikner}}, \bibinfo {author} {\bibfnamefont {J.~M.}\ \bibnamefont
  {Krueger}}, \ and\ \bibinfo {author} {\bibfnamefont {C.~J.}\ \bibnamefont
  {Foot}},\ }\href {\doibase 10.1103/PhysRevA.71.013604} {\bibfield  {journal}
  {\bibinfo  {journal} {Phys. Rev. A}\ }\textbf {\bibinfo {volume} {71}},\
  \bibinfo {pages} {013604} (\bibinfo {year} {2005})}\BibitemShut {NoStop}%
\bibitem [{\citenamefont {Garrido~Alzar}\ \emph {et~al.}(2006)\citenamefont
  {Garrido~Alzar}, \citenamefont {Perrin}, \citenamefont {Garraway},\ and\
  \citenamefont {Lorent}}]{Garrido2006}%
  \BibitemOpen
  \bibfield  {author} {\bibinfo {author} {\bibfnamefont {C.~L.}\ \bibnamefont
  {Garrido~Alzar}}, \bibinfo {author} {\bibfnamefont {H.}~\bibnamefont
  {Perrin}}, \bibinfo {author} {\bibfnamefont {B.~M.}\ \bibnamefont
  {Garraway}}, \ and\ \bibinfo {author} {\bibfnamefont {V.}~\bibnamefont
  {Lorent}},\ }\href {\doibase 10.1103/PhysRevA.74.053413} {\bibfield
  {journal} {\bibinfo  {journal} {Phys. Rev. A}\ }\textbf {\bibinfo {volume}
  {74}},\ \bibinfo {pages} {053413} (\bibinfo {year} {2006})}\BibitemShut
  {NoStop}%
\bibitem [{\citenamefont {Hofferberth}\ \emph {et~al.}(2006)\citenamefont
  {Hofferberth}, \citenamefont {Lesanovsky}, \citenamefont {Fischer},
  \citenamefont {Verdu},\ and\ \citenamefont {Schmiedmayer}}]{Hofferberth2006}%
  \BibitemOpen
  \bibfield  {author} {\bibinfo {author} {\bibfnamefont {S.}~\bibnamefont
  {Hofferberth}}, \bibinfo {author} {\bibfnamefont {I.}~\bibnamefont
  {Lesanovsky}}, \bibinfo {author} {\bibfnamefont {B.}~\bibnamefont {Fischer}},
  \bibinfo {author} {\bibfnamefont {J.}~\bibnamefont {Verdu}}, \ and\ \bibinfo
  {author} {\bibfnamefont {J.}~\bibnamefont {Schmiedmayer}},\ }\href
  {http://dx.doi.org/10.1038/nphys420} {\bibfield  {journal} {\bibinfo
  {journal} {Nature Phys.}\ }\textbf {\bibinfo {volume} {2}},\ \bibinfo {pages}
  {710} (\bibinfo {year} {2006})}\BibitemShut {NoStop}%
\bibitem [{\citenamefont {{Kollengode Easwaran}}\ \emph
  {et~al.}(2010)\citenamefont {{Kollengode Easwaran}}, \citenamefont
  {Longchambon}, \citenamefont {Pottie}, \citenamefont {Lorent}, \citenamefont
  {Perrin},\ and\ \citenamefont {Garraway}}]{KollengodeEaswaran2010}%
  \BibitemOpen
  \bibfield  {author} {\bibinfo {author} {\bibfnamefont {R.}~\bibnamefont
  {{Kollengode Easwaran}}}, \bibinfo {author} {\bibfnamefont {L.}~\bibnamefont
  {Longchambon}}, \bibinfo {author} {\bibfnamefont {P.-E.}\ \bibnamefont
  {Pottie}}, \bibinfo {author} {\bibfnamefont {V.}~\bibnamefont {Lorent}},
  \bibinfo {author} {\bibfnamefont {H.}~\bibnamefont {Perrin}}, \ and\ \bibinfo
  {author} {\bibfnamefont {B.~M.}\ \bibnamefont {Garraway}},\ }\href
  {http://stacks.iop.org/0953-4075/43/i=6/a=065302} {\bibfield  {journal}
  {\bibinfo  {journal} {Journal of Physics B: Atomic, Molecular and Optical
  Physics}\ }\textbf {\bibinfo {volume} {43}},\ \bibinfo {pages} {065302}
  (\bibinfo {year} {2010})}\BibitemShut {NoStop}%
\bibitem [{Note3()}]{Note3}%
  \BibitemOpen
  \bibinfo {note} {The antennas are made of ten loops of 0.71~mm diameter
  copper wire. They have a square shape with a 16~mm side, and are located at
  about 10~mm from the atoms.}\BibitemShut {Stop}%
\bibitem [{\citenamefont {Hofferberth}\ \emph {et~al.}(2007)\citenamefont
  {Hofferberth}, \citenamefont {Fischer}, \citenamefont {Schumm}, \citenamefont
  {Schmiedmayer},\ and\ \citenamefont {Lesanovsky}}]{Hofferberth2007}%
  \BibitemOpen
  \bibfield  {author} {\bibinfo {author} {\bibfnamefont {S.}~\bibnamefont
  {Hofferberth}}, \bibinfo {author} {\bibfnamefont {B.}~\bibnamefont
  {Fischer}}, \bibinfo {author} {\bibfnamefont {T.}~\bibnamefont {Schumm}},
  \bibinfo {author} {\bibfnamefont {J.}~\bibnamefont {Schmiedmayer}}, \ and\
  \bibinfo {author} {\bibfnamefont {I.}~\bibnamefont {Lesanovsky}},\ }\href
  {\doibase 10.1103/PhysRevA.76.013401} {\bibfield  {journal} {\bibinfo
  {journal} {Phys. Rev. A}\ }\textbf {\bibinfo {volume} {76}},\ \bibinfo
  {pages} {013401} (\bibinfo {year} {2007})}\BibitemShut {NoStop}%
\bibitem [{\citenamefont {van Es}\ \emph {et~al.}(2008)\citenamefont {van Es},
  \citenamefont {Whitlock}, \citenamefont {Fernholz}, \citenamefont {van
  Amerongen},\ and\ \citenamefont {van Druten}}]{vanEs2008}%
  \BibitemOpen
  \bibfield  {author} {\bibinfo {author} {\bibfnamefont {J.~J.~P.}\
  \bibnamefont {van Es}}, \bibinfo {author} {\bibfnamefont {S.}~\bibnamefont
  {Whitlock}}, \bibinfo {author} {\bibfnamefont {T.}~\bibnamefont {Fernholz}},
  \bibinfo {author} {\bibfnamefont {A.~H.}\ \bibnamefont {van Amerongen}}, \
  and\ \bibinfo {author} {\bibfnamefont {N.~J.}\ \bibnamefont {van Druten}},\
  }\href {\doibase 10.1103/PhysRevA.77.063623} {\bibfield  {journal} {\bibinfo
  {journal} {Phys. Rev. A}\ }\textbf {\bibinfo {volume} {77}},\ \bibinfo
  {pages} {063623} (\bibinfo {year} {2008})}\BibitemShut {NoStop}%
\bibitem [{\citenamefont {Morizot}\ \emph {et~al.}(2008)\citenamefont
  {Morizot}, \citenamefont {Longchambon}, \citenamefont {Kollengode~Easwaran},
  \citenamefont {Dubessy}, \citenamefont {Knyazchyan}, \citenamefont {Pottie},
  \citenamefont {Lorent},\ and\ \citenamefont {Perrin}}]{Morizot2008}%
  \BibitemOpen
  \bibfield  {author} {\bibinfo {author} {\bibfnamefont {O.}~\bibnamefont
  {Morizot}}, \bibinfo {author} {\bibfnamefont {L.}~\bibnamefont
  {Longchambon}}, \bibinfo {author} {\bibfnamefont {R.}~\bibnamefont
  {Kollengode~Easwaran}}, \bibinfo {author} {\bibfnamefont {R.}~\bibnamefont
  {Dubessy}}, \bibinfo {author} {\bibfnamefont {E.}~\bibnamefont {Knyazchyan}},
  \bibinfo {author} {\bibfnamefont {P.-E.}\ \bibnamefont {Pottie}}, \bibinfo
  {author} {\bibfnamefont {V.}~\bibnamefont {Lorent}}, \ and\ \bibinfo {author}
  {\bibfnamefont {H.}~\bibnamefont {Perrin}},\ }\href {\doibase
  10.1140/epjd/e2008-00050-2} {\bibfield  {journal} {\bibinfo  {journal} {Eur.
  Phys. J. D}\ }\textbf {\bibinfo {volume} {47}},\ \bibinfo {pages} {209}
  (\bibinfo {year} {2008})}\BibitemShut {NoStop}%
\bibitem [{\citenamefont {Zobay}\ and\ \citenamefont
  {Garraway}(2004)}]{Zobay2004}%
  \BibitemOpen
  \bibfield  {author} {\bibinfo {author} {\bibfnamefont {O.}~\bibnamefont
  {Zobay}}\ and\ \bibinfo {author} {\bibfnamefont {B.~M.}\ \bibnamefont
  {Garraway}},\ }\href {\doibase 10.1103/PhysRevA.69.023605} {\bibfield
  {journal} {\bibinfo  {journal} {Phys. Rev. A}\ }\textbf {\bibinfo {volume}
  {69}},\ \bibinfo {pages} {023605} (\bibinfo {year} {2004})}\BibitemShut
  {NoStop}%
\bibitem [{\citenamefont {Brink}\ and\ \citenamefont
  {Sukumar}(2006)}]{Brink2006}%
  \BibitemOpen
  \bibfield  {author} {\bibinfo {author} {\bibfnamefont {D.~M.}\ \bibnamefont
  {Brink}}\ and\ \bibinfo {author} {\bibfnamefont {C.~V.}\ \bibnamefont
  {Sukumar}},\ }\href {\doibase 10.1103/PhysRevA.74.035401} {\bibfield
  {journal} {\bibinfo  {journal} {Phys. Rev. A}\ }\textbf {\bibinfo {volume}
  {74}},\ \bibinfo {pages} {035401} (\bibinfo {year} {2006})}\BibitemShut
  {NoStop}%
\bibitem [{\citenamefont {Pitaevskii}\ and\ \citenamefont
  {Stringari}(2003)}]{PitaevskiiStringari}%
  \BibitemOpen
  \bibfield  {author} {\bibinfo {author} {\bibfnamefont {L.}~\bibnamefont
  {Pitaevskii}}\ and\ \bibinfo {author} {\bibfnamefont {S.}~\bibnamefont
  {Stringari}},\ }\href@noop {} {\emph {\bibinfo {title} {Bose-{E}instein
  Condensation}}}\ (\bibinfo  {publisher} {Oxford University Press},\ \bibinfo
  {year} {2003})\BibitemShut {NoStop}%
\bibitem [{\citenamefont {Marag\`o}\ \emph {et~al.}(2000)\citenamefont
  {Marag\`o}, \citenamefont {Hopkins}, \citenamefont {Arlt}, \citenamefont
  {Hodby}, \citenamefont {Hechenblaikner},\ and\ \citenamefont
  {Foot}}]{Marago2000}%
  \BibitemOpen
  \bibfield  {author} {\bibinfo {author} {\bibfnamefont {O.~M.}\ \bibnamefont
  {Marag\`o}}, \bibinfo {author} {\bibfnamefont {S.~A.}\ \bibnamefont
  {Hopkins}}, \bibinfo {author} {\bibfnamefont {J.}~\bibnamefont {Arlt}},
  \bibinfo {author} {\bibfnamefont {E.}~\bibnamefont {Hodby}}, \bibinfo
  {author} {\bibfnamefont {G.}~\bibnamefont {Hechenblaikner}}, \ and\ \bibinfo
  {author} {\bibfnamefont {C.~J.}\ \bibnamefont {Foot}},\ }\href {\doibase
  10.1103/PhysRevLett.84.2056} {\bibfield  {journal} {\bibinfo  {journal}
  {Phys. Rev. Lett.}\ }\textbf {\bibinfo {volume} {84}},\ \bibinfo {pages}
  {2056} (\bibinfo {year} {2000})}\BibitemShut {NoStop}%
\bibitem [{\citenamefont {Vogt}\ \emph {et~al.}(2012)\citenamefont {Vogt},
  \citenamefont {Feld}, \citenamefont {Fr\"ohlich}, \citenamefont {Pertot},
  \citenamefont {Koschorreck},\ and\ \citenamefont {K\"ohl}}]{Vogt2012}%
  \BibitemOpen
  \bibfield  {author} {\bibinfo {author} {\bibfnamefont {E.}~\bibnamefont
  {Vogt}}, \bibinfo {author} {\bibfnamefont {M.}~\bibnamefont {Feld}}, \bibinfo
  {author} {\bibfnamefont {B.}~\bibnamefont {Fr\"ohlich}}, \bibinfo {author}
  {\bibfnamefont {D.}~\bibnamefont {Pertot}}, \bibinfo {author} {\bibfnamefont
  {M.}~\bibnamefont {Koschorreck}}, \ and\ \bibinfo {author} {\bibfnamefont
  {M.}~\bibnamefont {K\"ohl}},\ }\href {\doibase
  10.1103/PhysRevLett.108.070404} {\bibfield  {journal} {\bibinfo  {journal}
  {Phys. Rev. Lett.}\ }\textbf {\bibinfo {volume} {108}},\ \bibinfo {pages}
  {070404} (\bibinfo {year} {2012})}\BibitemShut {NoStop}%
\bibitem [{\citenamefont {Bismut}\ \emph {et~al.}(2010)\citenamefont {Bismut},
  \citenamefont {Pasquiou}, \citenamefont {Mar\'echal}, \citenamefont {Pedri},
  \citenamefont {Vernac}, \citenamefont {Gorceix},\ and\ \citenamefont
  {Laburthe-Tolra}}]{Bismut2010}%
  \BibitemOpen
  \bibfield  {author} {\bibinfo {author} {\bibfnamefont {G.}~\bibnamefont
  {Bismut}}, \bibinfo {author} {\bibfnamefont {B.}~\bibnamefont {Pasquiou}},
  \bibinfo {author} {\bibfnamefont {E.}~\bibnamefont {Mar\'echal}}, \bibinfo
  {author} {\bibfnamefont {P.}~\bibnamefont {Pedri}}, \bibinfo {author}
  {\bibfnamefont {L.}~\bibnamefont {Vernac}}, \bibinfo {author} {\bibfnamefont
  {O.}~\bibnamefont {Gorceix}}, \ and\ \bibinfo {author} {\bibfnamefont
  {B.}~\bibnamefont {Laburthe-Tolra}},\ }\href {\doibase
  10.1103/PhysRevLett.105.040404} {\bibfield  {journal} {\bibinfo  {journal}
  {Phys. Rev. Lett.}\ }\textbf {\bibinfo {volume} {105}},\ \bibinfo {pages}
  {040404} (\bibinfo {year} {2010})}\BibitemShut {NoStop}%
\bibitem [{\citenamefont {Olshanii}\ \emph {et~al.}(2010)\citenamefont
  {Olshanii}, \citenamefont {Perrin},\ and\ \citenamefont
  {Lorent}}]{Olshanii2010}%
  \BibitemOpen
  \bibfield  {author} {\bibinfo {author} {\bibfnamefont {M.}~\bibnamefont
  {Olshanii}}, \bibinfo {author} {\bibfnamefont {H.}~\bibnamefont {Perrin}}, \
  and\ \bibinfo {author} {\bibfnamefont {V.}~\bibnamefont {Lorent}},\ }\href
  {\doibase 10.1103/PhysRevLett.105.095302} {\bibfield  {journal} {\bibinfo
  {journal} {Phys. Rev. Lett.}\ }\textbf {\bibinfo {volume} {105}},\ \bibinfo
  {pages} {095302} (\bibinfo {year} {2010})}\BibitemShut {NoStop}%
\bibitem [{\citenamefont {Stringari}(1998)}]{Stringari1998a}%
  \BibitemOpen
  \bibfield  {author} {\bibinfo {author} {\bibfnamefont {S.}~\bibnamefont
  {Stringari}},\ }\href {\doibase 10.1103/PhysRevA.58.2385} {\bibfield
  {journal} {\bibinfo  {journal} {Phys. Rev. A}\ }\textbf {\bibinfo {volume}
  {58}},\ \bibinfo {pages} {2385} (\bibinfo {year} {1998})}\BibitemShut
  {NoStop}%
\bibitem [{\citenamefont {Lin~Xia}\ \emph {et~al.}(2012)\citenamefont
  {Lin~Xia}, \citenamefont {Lobser},\ and\ \citenamefont {Cornell}}]{Xia2012}%
  \BibitemOpen
  \bibfield  {author} {\bibinfo {author} {\bibfnamefont {L.}~\bibnamefont
  {Lin~Xia}}, \bibinfo {author} {\bibfnamefont {D.}~\bibnamefont {Lobser}}, \
  and\ \bibinfo {author} {\bibfnamefont {E.}~\bibnamefont {Cornell}},\ }\href
  {http://meetings.aps.org/link/BAPS.2012.DAMOP.Q1.100} {\enquote {\bibinfo
  {title} {Collective excitations in quasi-2{D} condensates},}\ } (\bibinfo
  {year} {2012}),\ \bibinfo {note} {{DAMOP} poster session, poster
  {BAPS.2012.DAMOP.Q1.100}}\BibitemShut {NoStop}%
\bibitem [{\citenamefont {Olshanii}(2012)}]{Olshanii2012}%
  \BibitemOpen
  \bibfield  {author} {\bibinfo {author} {\bibfnamefont {M.}~\bibnamefont
  {Olshanii}},\ }\href@noop {} {} (\bibinfo {year} {2012}),\ \bibinfo {note}
  {private communication.}\BibitemShut {Stop}%
\bibitem [{\citenamefont {Castin}\ and\ \citenamefont
  {Dum}(1996)}]{Castin1996}%
  \BibitemOpen
  \bibfield  {author} {\bibinfo {author} {\bibfnamefont {Y.}~\bibnamefont
  {Castin}}\ and\ \bibinfo {author} {\bibfnamefont {R.}~\bibnamefont {Dum}},\
  }\href {\doibase 10.1103/PhysRevLett.77.5315} {\bibfield  {journal} {\bibinfo
   {journal} {Phys. Rev. Lett.}\ }\textbf {\bibinfo {volume} {77}},\ \bibinfo
  {pages} {5315} (\bibinfo {year} {1996})}\BibitemShut {NoStop}%
\bibitem [{\citenamefont {Salasnich}\ \emph {et~al.}(2004)\citenamefont
  {Salasnich}, \citenamefont {Parola},\ and\ \citenamefont
  {Reatto}}]{Salasnich2004}%
  \BibitemOpen
  \bibfield  {author} {\bibinfo {author} {\bibfnamefont {L.}~\bibnamefont
  {Salasnich}}, \bibinfo {author} {\bibfnamefont {A.}~\bibnamefont {Parola}}, \
  and\ \bibinfo {author} {\bibfnamefont {L.}~\bibnamefont {Reatto}},\ }\href
  {\doibase 10.1103/PhysRevA.69.045601} {\bibfield  {journal} {\bibinfo
  {journal} {Phys. Rev. A}\ }\textbf {\bibinfo {volume} {69}},\ \bibinfo
  {pages} {045601} (\bibinfo {year} {2004})}\BibitemShut {NoStop}%
\end{thebibliography}
%

\end{document}